\documentclass[manuscript]{emulateapj}

\slugcomment{Accepted for publication in \apj.}

\shorttitle{GCLF as distance indicator}
\shortauthors{Villegas et~al.}

\begin{document}

\title{The ACS Fornax Cluster Survey. VIII. The Luminosity Function of
Globular Clusters in Virgo and Fornax Early-Type Galaxies and its Use
as a Distance Indicator\altaffilmark{1}}

\author{Daniela Villegas\altaffilmark{2}, 
Andr\'es Jord\'an\altaffilmark{3,4}, 
Eric W. Peng\altaffilmark{5}, 
John P. Blakeslee\altaffilmark{6}, 
Patrick C\^ot\'e\altaffilmark{6}, 
Laura Ferrarese\altaffilmark{6}, 
Markus Kissler-Patig\altaffilmark{2}, 
Simona Mei\altaffilmark{7,8}, 
Leopoldo Infante\altaffilmark{3}, 
John L. Tonry\altaffilmark{9}, 
Michael J. West\altaffilmark{10} }

\altaffiltext{1}{Based on observations with the NASA/ESA Hubble Space
  Telescope, obtained at the Space Telescope Science Institute
  (STScI), which is operated by the Association of Universities for
  Research in Astronomy, Inc., under NASA contract \mbox{NAS
  5-26555}.}
\altaffiltext{2}{European Southern Observatory,
  Karl-Schwarzschild-Strasse 2, 85748 Garching bei M\"unchen, Germany}
\altaffiltext{3}{Departmento de Astronom\'ia y Astrof\'isica,
Pontificia Universidad Cat\'olica de Chile, Av.\
Vicu\~na Mackenna 4860, 7820436 Macul, Santiago, Chile
}
\altaffiltext{4}{Harvard-Smithsonian Center for Astrophysics, 60
  Garden St, Cambridge, MA 02138}
\altaffiltext{5}{Department of Astronomy, Peking University, Beijing
  100871, China}
\altaffiltext{6}{Herzberg Institute of Astrophysics, Victoria, BC V9E
  2E7, Canada.}
\altaffiltext{7}{University of Paris Denis Diderot,  75205 Paris Cedex
13, France}
\altaffiltext{8}{GEPI, Observatoire de Paris, Section de Meudon, 5 Place
J. Janssen, 92195 Meudon Cedex}
\altaffiltext{9}{Institute for Astronomy, University of Hawaii,
  Honolulu, HI 96822}
\altaffiltext{10}{European Southern Observatory, Alonso de C\'ordova
3107, Vitacura, Casilla 19001, Santiago, Chile}

\begin{abstract}
We use a highly homogeneous set of data from 132 early-type galaxies
in the Virgo and Fornax clusters in order to study the properties of
the globular cluster luminosity function (GCLF). The globular cluster
system of each galaxy was studied using a maximum likelihood approach
to model the intrinsic GCLF after accounting for contamination and
completeness effects. The results presented here update our Virgo
measurements and confirm our previous results showing a tight
correlation between the dispersion of the GCLF and the absolute
magnitude of the parent galaxy. Regarding the use of the GCLF as a
standard candle, we have found that the relative distance modulus
between the Virgo and Fornax clusters is systematically lower than the
one derived by other distance estimators, and in particular it is 
0.22\textsl{mag} lower than the value derived from surface brightness 
fluctuation measurements performed on the same data. From numerical 
simulations aimed at reproducing the observed dispersion of the value 
of the turnover magnitude in each galaxy cluster we estimate an intrinsic
dispersion on this parameter of 0.21\textsl{mag} and 0.15\textsl{mag} for 
Virgo and Fornax respectively. All in all, our study shows that the GCLF
properties vary systematically with galaxy mass showing no evidence 
for a dichotomy between giant and dwarf early-type galaxies. These properties
may be influenced by the cluster environment as suggested 
by cosmological simulations.

\end{abstract}


\section{Introduction}
\label{sec:INTRO}

The distribution of globular cluster (GC) magnitudes has the
remarkable property that it is observed to peak at a value of $M_V
\approx -7.5$ \textsl{mag} in a near universal fashion (e.g., Jacoby
et~al.~1992, Harris 2001, Brodie \& Strader 2006). This distribution,
usually referred to as the GC luminosity function (GCLF), has been
historically described by a Gaussian.  By virtue of its near
universality, the derived mean or ``turnover'' magnitude $\mu$ has
seen widespread use as a distance indicator (e.g.~Secker 1992, Sandage
\& Tammann 1995), even though some dispersion and discrepant results
have been reported in the literature (see discussion in Ferrarese
et~al.~2000a).

There is nevertheless no solid theoretical explanation
for the observed universality of the turnover magnitude. The
luminosity function is a reflection of the more fundamental mass
spectrum of the GCs, and as such the ``universal'' turnover magnitude 
corresponds to a cluster mass of $\sim 2 \times
10^5 \mathcal{M_{\odot}}$.
Vast efforts have been undertaken from the theoretical point of view
in order to explain the underlying universal mass function. The many
publications on this topic can be separated into those trying to
identify some particular initial condition that selects a certain mass
scale for star formation (e.g., Peebles \& Dicke 1968, Fall \& Rees
1985, West 1993), and those looking for a destruction mechanism that
selects clusters in a particular mass range starting from an initially
wide mass spectrum (e.g., Fall \& Rees 1977, Gnedin \& Ostriker 1997,
Prieto \& Gnedin 2008)

At the high-mass end (i.e.~$m_{gc} < \mu$) the mass function of
globular clusters resembles very closely the mass function of young
clusters and molecular clouds in the Milky Way and other nearby
galaxies (see e.g.~Harris \& Pudritz 1994, Elmegreen \& Efremov 1997,
Gieles et~al.~2006). On the other hand, neither young clusters nor
molecular clouds show a turnover on their mass distributions, but they
keep rising monotonically following a power-law to lower masses. 
Fall \& Zhang (2001) used simple analytical models (including 
evaporation by two-body relaxation, gravitational shocks and mass 
loss by stellar evolution) to study the evolution of the GC mass function.  
They showed that, for a wide variety of initial conditions, an initial 
power-law mass function develops a turnover that, after 12 Gyr, is 
remarkably close to the observed turnover of the GCLF. Vesperini 
(2000, 2001) reaches a similar conclusion, but finds that a log-normal
mass function provides a better fit to the data.
Fainter than the turnover, the evolution would
be dominated by two-body relaxation, and the mass function would end up 
having a constant number of GCs per unit mass, reflecting the
fact that the masses of tidally limited clusters are assumed to
decrease linearly with time until they are destroyed (other authors
propose different mass-loss rates, see e.g., Lamers et~al.~2006). 
Brighter than the turnover, the evolution is dominated by 
stellar evolution at early times and by gravitational shocks at late times.  
Recently, McLaughlin \& Fall (2008) have shown that the GC mass function 
in the Milky Way depends on cluster half-mass density (i.e.~the mean density 
within a radius containing half the total mass of the GC), in the sense that 
the turnover mass increases with half-mass density, while the width of the 
GC mass function decreases.  But while there is currently a fairly good
understanding of the dynamical processes that shape the GCLF, many
details are still missing. In particular none of the theories proposed has
been entirely successful on addressing the question of how the
turnover magnitude can remain constant regardless of environmental
properties and the mass of the host galaxy.
 
The use of deep HST data during the last years has resulted in high
quality GCLF data, reaching 2 magnitudes beyond the turnover at the
distance of the Virgo cluster ($\sim$16.5 Mpc, Mei et~al.~2007). The use
of these deeper observations has recently uncovered a strong
correlation between the GCLF dispersion and the absolute magnitude of
the parent galaxy (Jord\'an et~al.~2006,~2007b), demonstrating the
non-universality of this parameter and, as a consequence, of the GCLF
as a whole. Here we present a study of the GCLF of 132 early type
galaxies aimed to perform a precise test of the GCLF as a distance 
indicator by comparing the relative distance between the Virgo and 
Fornax clusters derived using the GCLF to the one derived using an 
analysis of surface brightness fluctuations (SBF, Tonry \& Schneider 1988)
based on the same data (Blakeslee et~al.~2009). Previous papers in the
this series have presented an introduction to the survey (Jord\'an
et~al.~2007a), the properties of the central surface brightness
profiles of early-type galaxies (C\^ot\'e et~al.~2007) and a catalog
of SBF distances and a precise measurement of the Virgo-Fornax
distance (Blakeslee et~al.~2009).

The organization of this paper is as follows. In \S\ref{sec:OBS&DATA} we
present a description of the observations and data reduction
procedures. In \S\ref{sec:GCLF} we describe the GCLF model fitting,
and in \S\ref{sec:sigma_M} we compare the properties of the fits to
previous results regarding the dispersion of the GCLF. Section
\S\ref{sec:dist} is focused on determining how universal the value of
the turnover magnitude is, while in \S\ref{sec:StCand} we look for a
better understanding of the external parameters that might affect this
value. Finally, in \S\ref{sec:CONC} we summarize our results
and the main conclusions of this paper.

\section{Data and GCLF ingredients}
\label{sec:OBS&DATA}

Each one of the 132 galaxies included in this study was observed with the
Advanced Camera for Surveys (ACS) during a single Hubble Space
Telescope (HST) orbit, as part of the ACS Virgo Cluster Survey
(ACSVCS) and the ACS Fornax Cluster Survey (ACSFCS). The goals and
main observational features of these two surveys are extensively
discussed in C\^ot\'e et~al.~(2004) and Jord\'an et~al.~(2007a),
respectively. We refer the interested reader to these publications for
further details.

The surveys targeted a total of 100 galaxies in the Virgo cluster and
43 galaxies in Fornax, and included observations in the F475W
($\approx$ Sloan $g$) and F850LP ($\approx$ Sloan $z$) passbands, with
exposure times of $\sim$750s and $\sim$1210s respectively. In what
follows we will refer to the F475W filter as ``$g$'' and to F850LP as
``$z$'', due to their close proximity to the corresponding Sloan
passbands.

Jord\'an et~al.~(2004) describes the pipeline implemented to
automate the reduction procedure and analysis of all images in both
surveys. The final output from this pipeline is a preliminary catalog
of GC candidates and expected contamination per galaxy, including
photometric and morphological properties, that are later used to
evaluate the probability $p_{GC}$ that a given object is a GC (see
Jord\'an et~al.~2009 for details). For the purposes of this study, and
as defined on previous ACSVCS and ACSFCS papers, we constructed the GC
candidate samples by selecting all sources that have $p_{GC} \geq$0.5.

Our catalog of GC candidates in a given galaxy differs from the
intrinsic GC population due to two effects: the existence of
contamination in the sample and the level of completeness of the
observations.

In order to quantify the average number of contaminants per field of
view we have used archival ACS imaging of 17 blank-high latitude
fields that have been observed in both the $g$ and $z$ bands, to the same
or deeper depth than our images. These control fields were processed
using the same pipeline implemented for the science data, and were
then used to build customized control fields, as if a given galaxy was
in front of it (the details of this process are explained in Peng et
al.~2006, were also a full list of the control fields used is available). 
For each of our target galaxies, the result is a catalog 
containing 17 different estimates of the expected foreground and background 
contamination. These are later used to obtain an average estimate of the 
contamination in the field of view of a given galaxy.

The completeness function needs to be built considering four parameters: 
the magnitude of the source ($m$), its size as measured by the projected half 
light radius ($r_{h}$), its color ($(g-z)_0$), and the surface brightness of 
the local background over which the object lies ($I_{b}$). 
The completeness function $f(m, r_{h}, (g-z)_{0}, I_{b})$ was obtained by 
performing simulations that added model GCs of different sizes 
($r_{h}= (1, 3, 6, 10)$ pc), 
colors ($(g-z)_0 = (0.7, 1.1, 1.5, 1.9)$ \textsl{mag}), and with King (1966) 
concentration parameter of $c=1.5$ to the images. 
Although the effect of the color of the clusters has not been considered in 
previous publications (e.g.~Peng et~al.~2006, Jord\'an et~al.~2007b), 
we have now established that it also has a small but measurable
effect over the expected completeness. Overall, roughly 6 million fake 
GCs were added for the completeness tests for each color, with equal 
fractions at each of the four sizes and avoiding physical overlaps with 
sources already present. These images were then reduced through 
exactly the same procedure used with the science data. The final output of 
the process is a four dimensional table that is used to evaluate $f$ given an 
arbitrary set of $(m, r_{h}, (g-z)_{0}, I_{b})$. The random 
uncertainty in the mean completeness curve is essentially zero,  
so the completeness limits at 90\% and 50\% are robust and can be 
determined with negligible error for a given population of objects.

This paper focuses on the study of the 89 early-type galaxies discussed
by Jord\'an et~al.(2007b) and all 43 galaxies of the ACSFCS. Our analysis is 
restricted to those galaxies that have more than five GC candidates and for 
which we were able to usefully constrain the GCLF parameters. 
These restrictions exclude 11 galaxies in the Virgo sample but none in Fornax.

\section{GCLF Model Fitting} 
\label{sec:GCLF}

\begin{figure}
\begin{center}
\includegraphics[width=8cm]{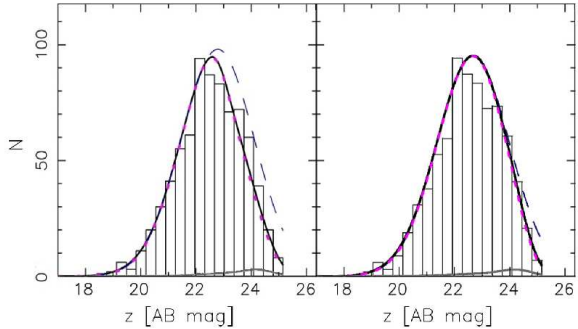}
\caption{Left: GCLF histogram for VCC1226 as presented in
Jord\'an et~al.~(2007b). The lines show the best-fit model (solid
black curve), the intrinsic Gaussian component (dashed curve), the
Gaussian component multiplied by the expected completeness (dotted
curve), and a kernel density estimate of the expected contamination
(solid gray curve). Right: The same as shown in the left hand side,
but now using the corrected completeness function.}
\label{fig:vcc1226_pXII}
\end{center}
\end{figure}

Given the observational information previously described we aim to
recover the parameters of the intrinsic luminosity function of the GCs
in a galaxy. We used a maximum likelihood approach similar to the one
described by Secker \& Harris (1993). According to this formalism, and
as detailed in Jord\'an et~al.~(2007b), we describe the intrinsic GCLF
by some function $G(m\!\mid\!\Theta)$, with $\Theta$ being the set of
model parameters to be fitted, and we assume that the uncertainties on
magnitude measurements $\epsilon_m$ have a Gaussian distribution. In
absence of contamination, the probability of observing a GC with a 
given effective radius $R_{h}$ and apparent magnitude $m$ against a 
galaxy background  $I_{b}$ would be:

\begin{equation}
\small G_{T}(m\!\mid\!\Theta,R_{h},I_{b},\epsilon_{m})= \mathcal{A}[ \
h(m\!\mid\!\epsilon_{m}) \, \otimes \, G(m\!\mid\!\Theta) \
]f(m,R_{h},I_{b}),
\end{equation}

\noindent where
$h(m\!\mid\!\epsilon_{m})=(2\pi\epsilon^{2}_{m})^{-1/2}
\mathrm{exp}(-m^2/2\epsilon^2_m)$, is the magnitude error
distribution, which is convolved with the intrinsic GCLF $G(m\!\mid\!\Theta)$. 
The normalization factor $\cal A$ is a function of the GCLF parameters
$\Theta$ and the GC properties $R_{h}$, $I_{b}$, and $\epsilon_m$, and
it is set by requiring that $G_{T}$ integrates to unity over the whole
magnitude range covered by the observations.

In practice a fraction $\mathcal{B}$ of the sources classified as GC
candidates in a galaxy are contaminants, so that the probability of
observing a \ GC \ with parameters $(m,R_{h},I_{b},\epsilon_{m})$ is
reduced by a factor $(1-\mathcal{B})$ and the distribution that
accounts for all the observed objects has to include the contaminants
luminosity function $b(m)$. Thus, the likelihood of observing a total
number of N objects with magnitudes $m_i$ and properties
$(R_{h},I_{b},\epsilon_{m})$ is

\begin{equation}
\mathcal{L}(\Theta, \mathcal{B})= \prod^{N}_{i=1} [(1-\mathcal{B})
G_{T}(m_{i}\!\mid\!\Theta,R_{h,i},I_{b,i},\epsilon_{m,i}) +
\mathcal{B}b(m_{i})],
\label{eq:likelihood}
\end{equation}

Jord\'an et~al.~(2007b) have made a detailed description of several
parametrization of the GCLF and their various advantages and
drawbacks. Here we focus on the study of the Gaussian representation,
because of its historic use in the study of the GCLF as a distance
indicator. It is worth noticing that other parametrization such a
$t_{5}$ function have also been successfully used for this purpose
(Secker 1992, Kissler et~al.~1994). For the case of a Gaussian the set
of model parameters will be $\Theta \equiv \{\mu,\sigma_{m}\}$, where
$\mu$ and $\sigma$ are the turnover and the dispersion in a
distribution of the form:

\begin{equation}
\frac{dN}{dm}=\frac{1}{\sqrt{2\pi}\sigma} \ \mathrm{exp}\left[-\frac{(m-\mu)^{2}}{2\sigma^{2}}\right]
\label{eq:gaussian}
\end{equation}

\begin{figure}
\begin{center}
\includegraphics[width=8.5cm]{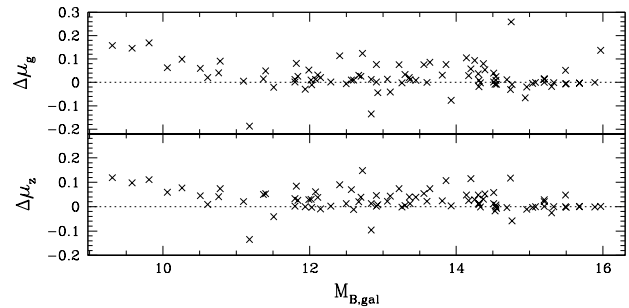}
\caption[Change in $\mu$ produced by using a corrected completeness
function.]{Difference in turnover magnitude produced by using the
completeness function presented by Jord\'an et~al. 2007b and the one
we are using here ($\Delta\mu \equiv \mu_{old} - \mu_{new}$) in the $g$ (top) and $z$ (bottom) bands,
vs. the $B$-band apparent magnitude of the parents galaxy.}
\label{fig:diff_mu}
\end{center}
\end{figure}

The coding implementation of the outlined maximum likelihood procedure
is in practice the same used to compute the GCLF by Jord\'an et
al.~(2007b)\footnote{In \S4.2 of Jord\'an et al.~(2007b) we show using
  simulations that our fitting procedures lead to no significant
  biases in the recovered $\mu$ and $\sigma$ for the range of GC
  system sizes in our sample.}, except that we are now using
completeness curves customized to the Fornax data too.
Also, during the analysis of the ACSFCS data we found a coding mistake in
the interpolation of the completeness curves previously used to
estimate the GCLF parameters of the Virgo galaxies.
The background information in the completeness curves was sometimes
misread in such a way that the completeness level assigned to a given
background brightness was lower than the real value.
As the changes in completeness are more significant for brighter 
backgrounds, massive galaxies were more affected than
dwarf galaxies. Even though it does not have any significant
effect over the main conclusions of Jord\'an et~al.~(2007b), we are
reporting the problem here because it produces a slight change in the
turnover magnitudes of the Virgo galaxies. The massive galaxies are
the most affected, with their turnover magnitudes becoming roughly
$\sim$0.1 \textsl{mag} brighter. This behavior can be observed in Figure
\ref{fig:vcc1226_pXII}, where we have plotted side-by-side the $z$-band
GCLF fit for VCC1226 as presented in Figure 4 of Jord\'an et
al.~(2007b), and the current fit implemented using the corrected
completeness function that now also includes a color correction. In
Figure \ref{fig:diff_mu} we have plotted the observed change in the
turnover magnitude ($\Delta \mu \equiv \mu_{old} - \mu_{new}$) in both bands, against the $B$-band
apparent magnitude of the parent galaxy, showing that the brightest
galaxies are the most evidently affected, unlike the dwarfs whose
turnover stays virtually unchanged. Some spread can be observed in the
case of the intermediate-luminosity galaxies, but in all cases the
change in $\mu$ is always lower than 0.15 \textsl{mag}.

\begin{center}
\begin{figure}
\includegraphics[scale=.99]{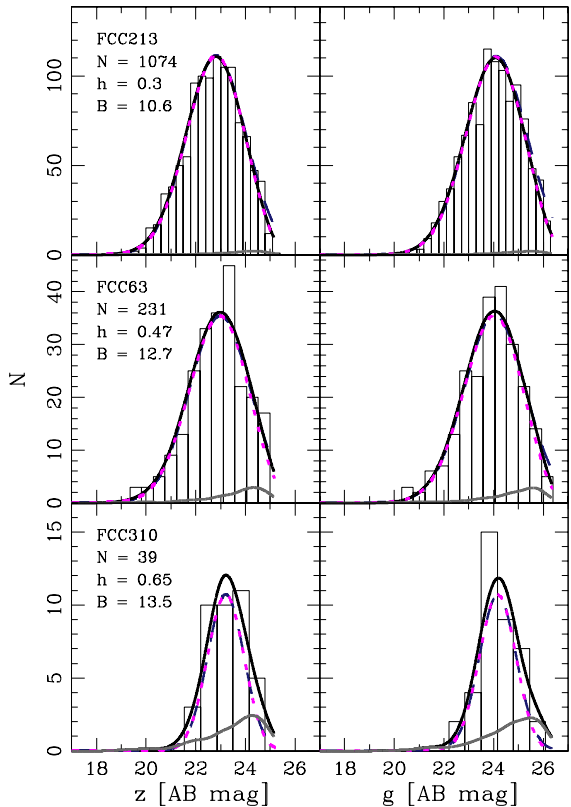}
\caption{GCLF histograms for the Virgo and Fornax sample galaxies. For
  each one of them we present the z- and $g$-band GCLFs side by
  side. The VCC/FCC name and $B$-band magnitude of the galaxy are
  indicated in the upper left corner of the left panel, where we also
  indicate the total number of sources $N$ in each histogram and the
  bin width $h$ used to construct it ($h$ is calculated as described
  in the text). In addition, we show the best-fit model (solid black
  curve), the intrinsic Gaussian component (dashed curve), the
  Gaussian component multiplied by the expected completeness (dotted
  curve), and a kernel density estimate of the expected contamination
  in the sample (solid gray curve). The solid black curve is the sum
  of the solid gray and dotted curves. The galaxies are ordered by
  decreasing apparent $B$-band total luminosity, reading down from the
  upper left corner. The parameters of the fits are given in Table
  \ref{tab:acsvcs_gclf_par} and \ref{tab:acsfcs_gclf_par}.  The full
  version of this figure is published in the electronic edition of the
  Astrophysical Journal. Sample panels are shown here for guidance
  regarding its form and content.}

\label{fig:gclf_plots}  
\end{figure}
\end{center}

Table~\ref{tab:acsvcs_gclf_par} lists the corrected values for the
Gaussian GCLF parameters of the ACSVCS galaxies. Updated values for
the evolved-Schechter function fits presented by Jord\'an et
al.~(2007b) will be presented elsewhere. The Gaussian parameters shown
in Table ~\ref{tab:acsvcs_gclf_par} are the ones considered for this
publication and they should be used for future reference. This table
includes, for all the ACSVCS galaxies: the $B$-band apparent magnitude
from Binggeli et~al.~(1985), the estimated GCLF parameters in both bands, 
the fraction of objects that are considered to be contaminants, and the 
total number of globular cluster candidates (including
contaminants). Table~\ref{tab:acsfcs_gclf_par} presents the equivalent
information computed for the ACSFCS galaxies, including the $B$-band
absolute magnitude from Ferguson (1989a). Figure \ref{fig:gclf_plots} 
shows the $z$ and $g$-band GCLF histograms of the sample galaxies, 
ordered by decreasing apparent $B$-band total luminosity. The dashed 
curve corresponds to the intrinsic Gaussian component given by 
Equation \ref{eq:gaussian} and the parameters in Table
\ref{tab:acsfcs_gclf_par}. The Gaussian component multiplied by the
expected completeness is represented by the dotted curve, and a kernel
density estimate of the expected contamination in the sample appears
as a solid gray curve. The solid black curve is the sum of the solid
gray and dotted curves, and corresponds to the net distribution for
which the likelihood in Equation \ref{eq:likelihood} is maximized. The
name and apparent B magnitude of the galaxy are indicated in the upper
left corner of the left panel, where we also quote the total number of
sources in each histogram and the bin width {\it h}. The width of the 
bins, used only for display purposes here, follows the rule 
$h = 2(\textsc{IQR})N^{-1/3}$ , where $(\textsc{IQR)}$ is the interquartile range of the 
magnitude distribution and $N$ is the total number of objects in each 
GC sample (Izenman, 1991).

\begin{figure}
\centering
\includegraphics[scale=0.45]{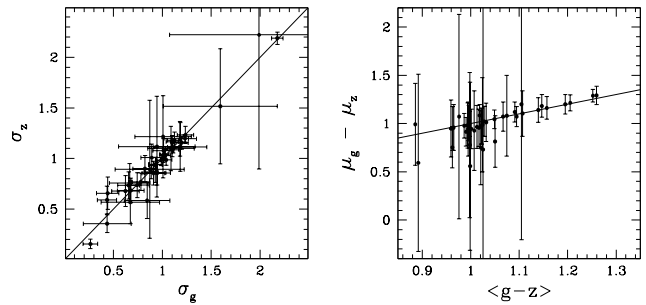}
\caption{Left: Estimate of Gaussian dispersion in the $z$ band,
$\sigma_{z}$, vs. the same quantity in the $g$ band, $\sigma_{g}$ , for
the GCLFs of our Fornax sample. Uncertainties are 1$\sigma$. The
line marks the one-to-one correspondence between these two
quantities. Right: Difference between estimates of Gaussian means in
the $g$ and $z$ bands, $\mu_g - \mu_z$ , vs. the mean color $\langle g - z
\rangle$ of the GC systems of our sample galaxies. Uncertainties are
1$\sigma$. The line marks the one-to-one correspondence between these
two quantities.}
\label{fig:one-by-one}
\end{figure}

As a sanity check of our fitting procedure, in the left-hand side of
Figure \ref{fig:one-by-one} we compare the Gaussian dispersion
inferred from the GCLF fit in each band, $\sigma_{g}$
vs. $\sigma_{z}$, including only data from the Fornax sample. In the right-hand
side of the same figure we have plotted the difference between
estimates of Gaussian means in the $g$ and $z$ bands $(\mu_g - \mu_z)$,
vs. the mean color $\langle g - z \rangle$ of the GC systems of our
sample galaxies. From the very tight correlation between the
measurements in different bands we conclude that the GCLF fitting
procedure is internally consistent and also that our error estimations
are realistic.

\section{The $\sigma - M_{B,\lowercase{gal}}$ Relation}
\label{sec:sigma_M}
\begin{figure}
\begin{center}
\includegraphics*[width=10cm]{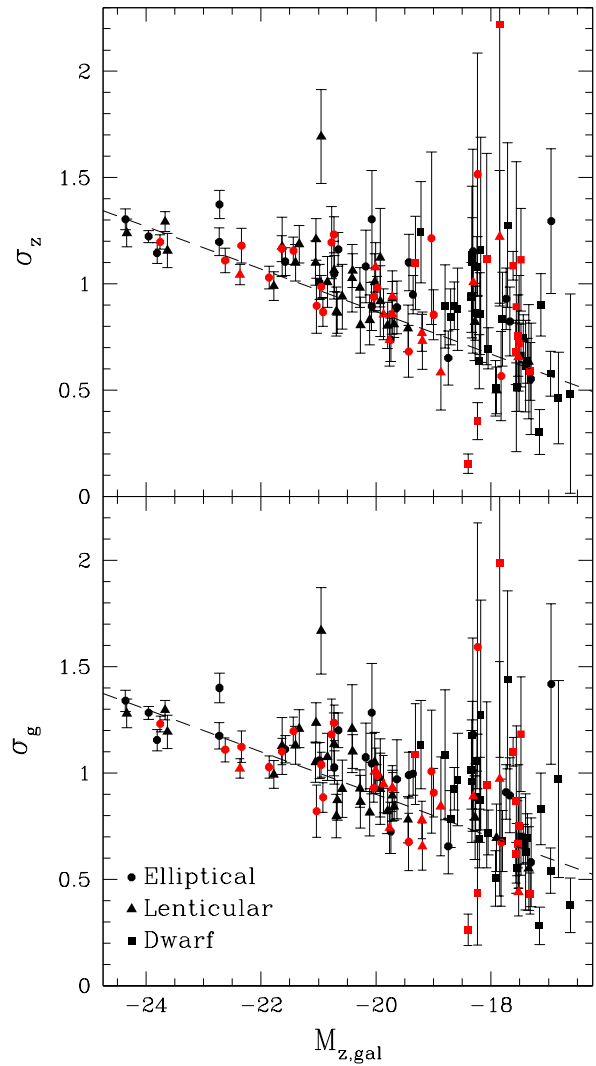}
\caption{Top: GCLF dispersion $\sigma_{z}$, inferred from Gaussian
fits to the $z$-band data, vs. galaxy $M_{z,gal}$. The dashed line
corresponds the linear relation between $\sigma_{z}$ and $M_{z,gal}$
in eq.~\ref{eq:sigma_z}. Bottom: Same comparison, but for the Gaussian
dispersion of the $g$-band GCLFs, $\sigma_{g}$. The dashed line
represents eq.~\ref{eq:sigma_g}. In both panels the black symbols
correspond to Virgo galaxies, and the red ones to the sample in the
Fornax cluster. We have morphologically separated the galaxies into
elliptical (circles), lenticular (triangles), and dwarf (squares)
galaxies.}
\label{sigma-M}
\end{center}
\end{figure}

One of the main results discussed in Jord\'an et~al.~(2006, 2007b) is
the existence of a strong correlation between the dispersion of the
GCLF $\sigma$, and the $B$-band absolute magnitude of the host galaxy
$M_{B,gal}$, with brighter galaxies showing higher dispersion
values. Even though some suggestive evidence on this respect was
previously presented by other authors (e.g.~Kundu \& Whitmore, 2001)
the high precision and homogeneity of our ACS/HST data unveiled the
$\sigma- M_{B,gal}$ correlation as a general trend in GC systems, which 
was later extended to still higher galaxy luminosity by Harris et~al.~(2009) 
using 5 giant elliptical galaxies in the Coma cluster.
Figure \ref{sigma-M} shows this correlation for all the 132 galaxies
in our sample in both bands, now using the homogeneous $z$-band absolute
magnitudes derived from the apparent magnitudes estimated by Ferrarese
et~al.~(2006) and C\^ot\'e et~al.~(2010, in preparation) and the corresponding
distance moduli published by Blakeslee et~al.~(2009). These values 
were corrected for reddening assuming $A_{z} = 1.485 \ E(B-V)$ 
(Ferrarese et~al.~2006) where the value of $E(B-V)$ was taken from 
Schlegel et~al.~(1998). In this figure we have used different symbols 
in order to identify the galaxies according to their morphological 
classification, but no particular trend related to this property seems 
to be obvious. The straight lines drawn in the panels correspond to 
error-weighted linear characterizations of these trends:

\begin{equation}
\sigma_{z}=(1.07 \pm 0.02) - (0.10 \pm 0.01) (M_{z,gal}+22)
\label{eq:sigma_z}
\end{equation}

\noindent and 

\begin{equation}
\sigma_{g}=(1.10 \pm 0.01) - (0.10 \pm 0.01) (M_{z,gal}+22).\\
\label{eq:sigma_g}
\end{equation}

We have excluded from these fits three galaxies for which no z-band 
magnitudes are available: VCC1535, VCC1030, and FCC167. Although shown 
in Figure \ref{sigma-M}, FCC21 (=NGC1316) is also not included on the fits
because the observed GC system in this galaxy is highly influenced by 
interaction and proximity with its satellite galaxies, and therefore our 
GCLF fit is not reliable. Unlike Jord\'an et~al.~(2006, 2007b) we have now 
also excluded from the analysis four galaxies in the Virgo cluster (VCC1297, 
VCC1199, VCC1192 and VCC1327) and two galaxies in Fornax (FCC202 and FCC143) 
because their GC systems appear to be contaminated by their proximity to 
massive ellipticals. All these galaxies are nonetheless retained in the 
Tables for completeness.

Equations \ref{eq:sigma_z} and \ref{eq:sigma_g} confirm the trend 
previously observed in Virgo, and with higher statistical significance, 
by including the Fornax data. This result shows by itself that the GCLF 
parameters are not universal and depend at least on one parameter,
i.e.~the luminosity of the parent galaxy, adding an additional feature 
that needs to be accounted for by theories aiming to explain the shape 
of the GCLF.

When the data corresponding to each cluster are fitted independently
the linear characterizations obtained are, in the case of Virgo:\\
\begin{eqnarray}
\sigma_{z}=(1.09 \pm 0.01) - (0.08 \pm 0.01) (M_{z,gal}+22), \\
\noindent \sigma_{g}=(1.11 \pm 0.01) - (0.09 \pm 0.01) (M_{z,gal}+22); 
\end{eqnarray}

\noindent and for the Fornax cluster: \\

\begin{eqnarray}
\sigma_{z}=(1.07 \pm 0.04) - (0.13 \pm 0.02) (M_{z,gal}+22), \\
\noindent \sigma_{g}=(1.09 \pm 0.03) - (0.10 \pm 0.01) (M_{z,gal}+22).  
\end{eqnarray}

\noindent This translates into a $0.05-0.1$ \textsl{mag} difference in 
dispersion at $M_{z} \sim 22$, and also shows that the linear fits derived 
from both sets of data are equivalent within the uncertainties.

As discussed in Jord\'an et~al.~(2006) it is rather straightforward to
link this trend in luminosity dispersion with a similar trend in the
mass distribution of GCs. It is well known that giant galaxies tend on
average to have more metal-rich GC populations when compared to
dwarfs, showing also larger dispersions in metallicity (see e.g.~Peng
et~al.~2006). This result, added to the dependence of the cluster
mass-to-light ratios ($\Upsilon$) on metallicity, opens the
possibility that the observed dispersion in the value of $\sigma$
might be metallicity-driven. These variations in $\Upsilon$ have a
strong dependence on wavelength. In bluer filters (the $g$-band in
our case), variations of a factor of 2 or more in $\Upsilon$ can be
observed in the typical metallicity range of GCs ($-2 \leq$ [Fe/H]
$\leq 0$). At redder wavelengths this variation becomes less dramatic,
as shown by old stellar population models (e.g.~PEGASE population
synthesis models; Fioc \& Rocca-Volmerange, 1997). In particular the
expected variation in $\sigma$ as a consequence of changes of
$\Upsilon$ in our $z$-band measurements should not be higher than
$\sim$ 4\%, which means that the spread in the value of $\sigma$
observed in the upper panel of Figure \ref{sigma-M} reflects almost
entirely a trend in the mass distribution of globular
clusters. Moreover, the very similar values obtained for $\sigma$ in
the $z$- and $g$-bands shows immediately that the trend of $\sigma$
with $M_B$ cannot be generated by metallicity-driven changes in $\Upsilon$.

\section{A Relative Virgo-Fornax Distance Estimation}
\label{sec:dist}
Several methods have been used in order to obtain accurate distance
estimations for both the Virgo and Fornax clusters, a task that is in
general more easily achieved in the case of Fornax due to its more
compact nature. The Virgo cluster extends for over 100 deg$^2$ in the
sky, showing a complex and irregular structure, with galaxies of
different morphological type showing different spatial and kinematic
distributions. Working on these conditions the various distance
estimators have reached different levels of accuracy (see Ferrarese et
al. 2000a, 2000b). We will discuss now a compilation of results from
the literature, which are also summarized in Table \ref{tab:dists}.

The HST Key Project to measure the Hubble constant aimed at obtaining 
accurate distances to galaxies using the period-luminosity relation for 
Cepheid variables (their final results are presented in Freedman et al.~2001).
It included the identification of Cepheids belonging to 6 spiral
galaxies in Virgo and 2 in Fornax, that were used to estimate
the distance to their parent galaxies, and then to the corresponding
clusters. This resulted in distance moduli of \mbox{$(m-M)_{V}=30.92 \pm
0.05$ \textsl{mag}} and \mbox{$(m-M)_{F}=31.39 \pm 0.20$ \textsl{mag}}
for Virgo and Fornax respectively, which translates into a relative 
distance modulus of $\Delta(m-M) = 0.47 \pm 0.20$ \textsl{mag}.

D'Onofrio et~al.~(1997) derived the relative distance between Virgo
and Fornax by applying the \mbox{D$_{n}-\sigma$} (Dressler et~al.,
1987) and the fundamental plane (Djorgovski \& Davis, 1987) relations
to a homogeneous sample of early-type galaxies.
The two distance indicators gave consistent results with
a relative distance modulus of \mbox{$\Delta(m-M) = 0.45 \pm
0.15$ \textsl{mag}}. These results are in close agreement with the value
\mbox{$\Delta(m-M) = 0.52 \pm 0.17$ \textsl{mag}} later published by Kelson et
al.~(2000) obtained also by using the fundamental plane and
\mbox{D$_{n}-\sigma$} relations built from data calibrated by the
aforementioned Cepheid distances to spiral galaxies in both Virgo and
Fornax.

The planetary nebula luminosity function (PNLF) has also been used for
measuring distances in the local Universe. Ciardullo et~al.~(1998)
determined a distance modulus of \mbox{$(m-M)_{V}=30.79 \pm 0.16$ \textsl{mag}}
to M87, in good agreement with previous measurements (e.g.~Jacoby et~al.,
1990). McMillan et~al.~(1993) used the PNLF to determine the distance 
to three galaxies in Fornax, obtaining a mean distance to the cluster 
of \mbox{$(m-M)_{F}= 31.14 \pm 0.14$ \textsl{mag}}. If we consider M87 to be 
at the center of Virgo, the corresponding relative distance modulus would be
\mbox{$\Delta(m-M) = 0.35 \pm 0.21$ \textsl{mag}}. Ferrarese et~al.~(2000a) 
calibrated literature measurements of the PNLF using Cepheids, which led 
them to estimate a relative distance modulus between Virgo and Fornax of 
\mbox{$\Delta(m-M) = 0.30 \pm 0.10$ \textsl{mag}} when considering the 
A-subcluster as indicative of the distance to Virgo.

Earlier relative distance modulus results derived by using the GCLF as
distance indicator present some hints of disagreement with the other 
estimations discussed here. Even though they were working with small 
and rather heterogeneous samples, previous studies tend to put this value
around a very low \mbox{$\Delta(m-M) \sim 0.13$ \textsl{mag}} 
(e.g.~Kohle et~al.~1996, Blakeslee \& Tonry 1996, Ferrarese et~al.~2000a, 
Richtler 2003).

One of the most reliable distance estimators when it comes to
population II samples is the surface brightness fluctuations (SBF)
method due to its high internal precision. The ACS Virgo and Fornax
clusters surveys, among whose aims is studying GC
properties and measuring surface brightness fluctuations,
provide us with the ideal data for comparing the properties of the
GCLF as a distance estimator with SBF results. We will discuss these 
results separately in the next session.

\subsection{SBF distances}
The method of SBF was first introduced by Tonry \& Schneider (1988),
and uses the fluctuations produced in each pixel of an image by the
Poissonian distribution of unresolved stars in a galaxy in order to
estimate the distance to the object.  The amplitude of those surface
brightness fluctuations normalized to the underlying mean galaxy
luminosity are inversely proportional to distance and can therefore be
used as a distance indicator (see Blakeslee et~al.~1999 for a review).

The distances to the Virgo galaxies included in the ACSVCS have been
measured using the SBF method. Mei et~al.~(2005a) describes the
reduction procedure used for the surface brightness analysis of the
ACSVCS data, and Mei et~al.~(2005b) presents the calibration for giant
and dwarf early-type galaxies. Finally, Mei et~al.~(2007) introduces
the distance catalog for a total 84 galaxies (50 giants and 34 dwarf)
for which the SBF method was successfully implemented, delivering at
the same time a three dimensional map of the structure of the Virgo
cluster. These distance values were later updated and the measurements
extended to include the 43 early-type galaxies of the ACSFCS in 
Blakeslee et~al.~(2009). In our analysis we will use the consistent
set of Virgo and Fornax distances presented by the later publication.
When no SBF distance is available for one of our sample galaxies, we 
assume it is located at the mean Virgo distance ($(m-M)=31.09$~\textsl{mag}) 
adopted by Mei et~al.~(2007). This estimate is based on ground-based I-band 
SBF measurements calibrated against Cepheids distances (Tonry et~al.~2000,
Freedman et~al.~2001). 

From their SBF measurements  Blakeslee et~al.~(2009) derives a 
relative Virgo-Fornax distance modulus of 
\mbox{$\Delta(m-M) = (0.42 \pm 0.03)$~\textsl{mag}}, which locates 
the Fornax cluster at a distance of \mbox{$d_{F}=20$~Mpc} 
(\mbox{$(m-M)_{F}=31.5$~\textsl{mag}}). This value is in good 
agreement with the relative distance moduli derived from the other 
distance estimators discussed above and summarized in Table 
\ref{tab:dists}, but it is significantly more precise.

\subsection{$\mu_{z}$ as Distance Indicator}
\label{sec:dist_in}
One of the main problems in understanding the properties of the
turnover of the GCLF as distance indicator is the lack of homogeneity
in the data. The most comprehensive compilation of recent data (mainly
HST observations) was presented by Richtler (2003), including a total
of 102 turnover magnitudes coming from at least 8 different
publications. This inhomogeneity introduces a major source of
uncertainty in the analysis, as one has to rely on each author's
results irrespective of the fact that they might not be using the same
procedure to reduce the data, the observations might not be on the
same photometric band, and they might not even be using the
same analytic form to fit the GCLF.

\begin{center}
\begin{figure}
\includegraphics*[width=8.2cm]{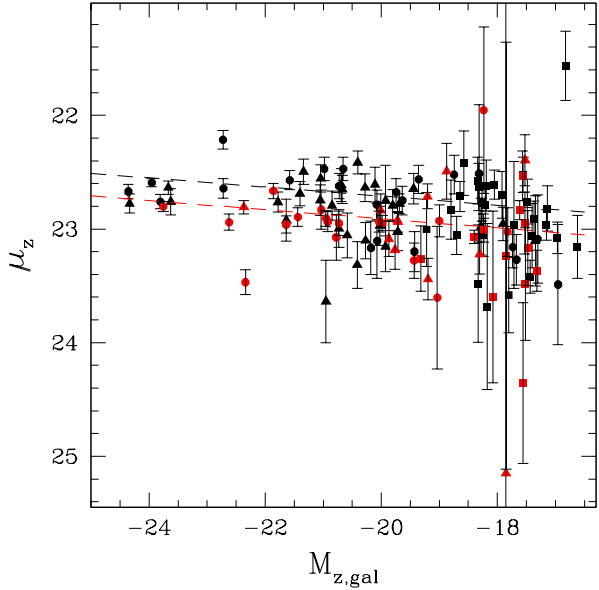}
\caption{GCLF turnover magnitude, $\mu_{z}$, inferred from Gaussian
fits to the $z$-band data, vs. $z$-band galaxy absolute magnitude
$M_{z,gal}$, for all the galaxies in the ACS Virgo (black symbols) and
Fornax (red symbols) cluster surveys. The lines represent a
simultaneous error-weighted linear fit performed over both samples,
that corresponds to $\mu_z = (23.51 \pm 0.11) + (0.04 \pm
0.01)M_{z,gal}$, plus and offset of \mbox{$\Delta (m-M)= 0.20 \pm 0.04$ 
\textsl{mag}} for the galaxies in Fornax. The sample has been
morphologically separated into elliptical (circles), lenticular
(triangles), and dwarf (squares) galaxies.\\}
\label{fig:tover_err}
\end{figure}
\end{center}

The data we are presenting here are the largest and most homogeneous
set of GCLF fits available to date. Our photometry is also deep enough
to cover the GCLF at least 2 magnitudes past the turnover, therefore
we are able to obtain reliable estimates of this parameter. In Figure
\ref{fig:tover_err} we have plotted the GCLF turnover magnitude  
against the $z$-band absolute magnitude of the parent galaxy. 
The lines show the best linear fit $p(M_{z,gal})$ to each cluster's
data, derived by minimizing the value of $\chi^2$ calculated as:

\begin{eqnarray}
\nonumber
\chi^2 &=& \sum_{i=1}^{n_{V}}  \left[ \frac{\mu^i_{z,V} - 
        p(M^i_{z,gal,V})}{\delta(\mu^i_{z,V})} \right]^2  \\
     & & + \sum_{j=1}^{n_{F}} \left[ \frac{(\mu^j_{z,F} - \Delta) - 
        p(M^j_{z,gal,F} - \Delta)}{\delta(\mu^j_{z,F})} \right]^2
\label{eq:chi_sq}
\end{eqnarray}

\noindent where the two sums are over the $n_V$ and $n_F$ galaxies in
Virgo and Fornax respectively, $\delta(\mu_{z})$ is the estimated 
error in $\mu_{z}$, and the offset $\Delta = \Delta(m-M)$ corresponds 
to the relative distance modulus. In this equation each one of the 
$M_{z,gal}$ components was estimated as:

\begin{eqnarray}
M^i_{z,gal} = m^i_{z,gal} - (m-M)_V - A^i_z
\label{eq:abs_mag}
\end{eqnarray}

\noindent where $(m-M)_V$=31.09 \textsl{mag} is the assumed mean 
distance modulus to the Virgo cluster. The four galaxies 
belonging to the W' cloud in our Virgo sample (VCC538, VCC571, VCC575, 
VCC731 and VCC 1025) were excluded from all our distance estimation 
fits as they are know to be located much further ($D \sim 23$ Mpc) 
than the mean Virgo distance. 

We have found that the best fit model for Eq.~\ref{eq:chi_sq}
corresponds to $\mu_z = (23.51 \pm 0.11) + (0.04 \pm 0.01)M_{z,gal}$
for a value of \mbox{$\Delta (m-M)$= 0.20 $\pm$ 0.04 \textsl{mag}},
where the error was estimated using bootstrap resampling of the data. 
This relative distance modulus represents a factor $\sim$2 difference 
with the results coming from most of the distance estimators previously 
described, and in particular it is $\sim$0.22 mag lower than the 
$\Delta (m-M)$= 0.42 \textsl{mag} derived by using the SBF method with 
the same data. It is important to stress that this discrepancy cannot 
be attributed to the data itself, because we are now using a large sample 
of highly homogeneous data. Also the fact that the $z$-band absolute 
magnitudes of the galaxies in both samples were derived from equivalent
observations and performing essentially the same analysis, minimizes
the amount of possible biases.

\begin{figure}
\includegraphics[width=8.6cm]{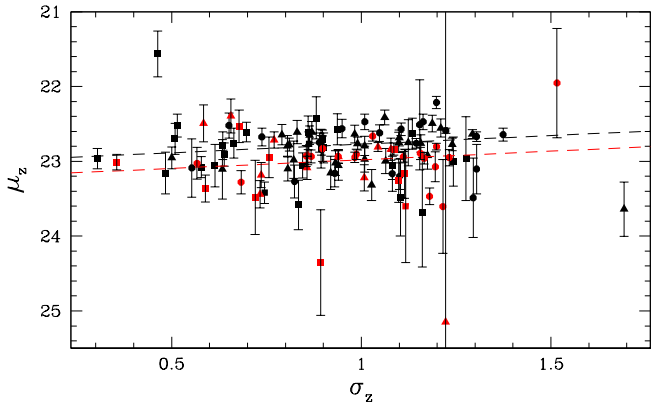}
\caption{GCLF turnover magnitude, $\mu_{z}$, vs. GCLF dispersion,
$\sigma_{z}$, both inferred from Gaussian fits to the $z$-band data, for
all the galaxies in the ACS Virgo (black symbols) and Fornax (red
symbols) cluster surveys. The sample has been morphologically
separated into elliptical (circles), lenticular (triangles), and dwarf
(squares) galaxies. The lines represent a simultaneous error-weighted
linear fit to the data corresponding to \mbox{$\mu_{z} = (22.99 \pm
0.04) - (0.23 \pm 0.04)\sigma_{z}$}, with an offset of $\Delta
\mu_{z}=0.21 \pm 0.04$ for the Fornax data. }
\label{fig:sigma_err}
\end{figure}

On the other hand, we are aiming to establish the level of precision
at which $\mu$ might be useful as a distance indicator and therefore
it seems natural to calibrate it against a parameter that is distance
independent, which is not the case for $M_{z,gal}$. The GCLF
dispersion, $\sigma$, appears like a good choice due to the already
established correlation between $\sigma$ and M$_{z,gal}$. In Figure
\ref{fig:sigma_err} we have plotted $\mu_{z}$ against $\sigma_{z}$ for
the complete sample in Virgo (black) and Fornax (red), separating the
galaxies by morphological type. A $\chi^2$ minimization equivalent to
Equation \ref{eq:chi_sq} was also performed in this case, obtaining as 
the best fit model: 
\mbox{$\mu_{z} = (22.99 \pm 0.04) - (0.23 \pm 0.04)\sigma_{z}$}. 
In this case the offset between both samples
corresponds to $\Delta (m-M)$= 0.21 $\pm$ 0.04 \textsl{mag}, where the 
error was estimated performing a bootstrap resampling of the data. This 
independent fit delivers a relative distance modulus that is
consistent with the previously derived value.

The observed difference between SBF and GCLF distances has already been
reported by Richtler (2003), attributing this phenomena to the
presence of intermediate-age GCs, which might contaminate the
sample. Our sample is made up exclusively of early type galaxies,
which are old stellar systems where the presence of intermediate-age
clusters is rarely observed (although some cases have been reported in
the literature, see e.g.~Goudfrooij et~al.~2001 for the case of
NGC1316 = FCC21, and Puzia et~al.~2002 for NGC4365 = VCC731), so it is 
unlikely that this is the reason of the observed
discrepancy. Ferrarese et~al.~(2000a) have consistently reported discrepancies
between their GCLF estimated distances and those obtained from other 
estimators (particularly SBF and PNLF). They found the GCLF turnover in 
Fornax to be a full 0.5$\textsl{mag}$ brighter than the value observed in 
Virgo. The internal errors in the GCLF measurements and the expected 
uncertainty due to cluster depth effects were not found to be enough to 
explain the scatter in their observations, suggesting the existence of 
a second parameter driving the GCLF turnover magnitude.

One obvious way to explain the observed discrepancy between GCLF and
SBF measurements would be a mean age difference for the Virgo and
Fornax cluster galaxies (i.e., the Fornax cluster galaxies might be
younger by some amount, leading to a brighter turnover).  The key
question, then, is determining the age difference that would be needed
to explain the observed $\sim$0.23 \textsl{mag} difference. According to the
Bruzual \& Charlot (2003) models, for a metallicity of $Z=0.004$ and a
Salpeter (1955) initial mass function, the observed offset would be
consistent with an age of roughly 9 Gyr for the Fornax cluster when
arbitrarily assuming an age of 12 Gyr for Virgo. This age difference
would also translate into slightly bluer mean colors for the Fornax
GCs, which should be on average $\sim$0.04 \textsl{mag} bluer than their Virgo
counterparts at a fixed galaxy mass. Performing a linear fit to the
GCs mean color $\langle g-z \rangle$ vs. M$_{z,gal}$ correlation of
our data we found that both clusters could follow the same trend but
including an offset of $0.022 \pm 0.015$ \textsl{mag} to redder colors in the
case of Virgo. Although almost consistent with zero, this value is also
consistent with the expected color discrepancy given by the necessary
age difference. The SBF technique, in which the fluctuations are 
calibrated against a measure of the stellar populations (i.e., color), 
would have this difference, if real, accounted for.

\subsection{The Observed Dispersion on the Value of $\mu_{z}$}
\begin{figure}
\includegraphics[width=8cm]{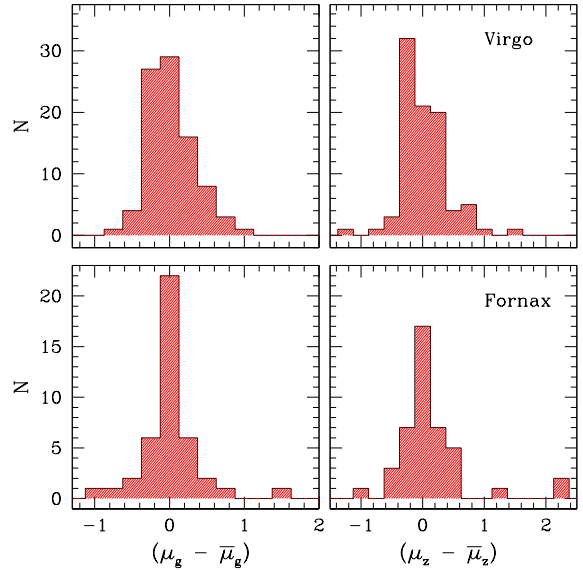}
\caption{Spread in magnitude around the weighted mean turnover
magnitude, for Virgo (top) and Fornax (bottom) in the $g$ (left) and z
(right) bands.}
\label{fig:MTO}
\end{figure}

A relatively large scatter can be observed in the turnover magnitude
values displayed in Figure \ref{fig:tover_err}. In this subsection we
want to address the question of how much of this dispersion is
intrinsic to the sample and how much is the result of observational
effects. The histograms in Figure \ref{fig:MTO} give a better
illustration of this scatter, where we have plotted the distribution
of magnitudes around the mean turnover magnitude of each sample, 
estimated though a 3-sigma clipping algorithm. Subtracting the
mean turnover magnitudes of both samples we obtain:
$(\bar{\mu}^g_F - \bar{\mu}^g_V)= 0.15$ \textsl{mag} and 
$(\bar{\mu}^z_F - \bar{\mu}^z_V)= 0.17$ \textsl{mag}, which delivers a 
first-order estimate of the relative Virgo-Fornax distance modulus.
We estimate the observed dispersion on the right ($z$-band) 
panels of Figure \ref{fig:MTO} in 0.31 \textsl{mag} and 0.28 \textsl{mag}, 
for Virgo and Fornax respectively, also using a 3-sigma clipping 
algorithm. We are more interested on studying 
the dispersion on the $z$-band because is much less sensitive to 
metallicity variations than the $g$-band.

There are then three main factors driving the spread: cluster depth, 
measurement errors and the intrinsic scatter in the turnover magnitude. 
From their 3D map of the Virgo cluster, Mei et~al.~(2007) have determined 
that the back-to-front depth of the cluster measured from our sample of
galaxies is 2.4$\pm$0.4 Mpc ($\pm 2\sigma$ of the intrinsic distance
distribution). At the Virgo distance this translates into a dispersion
due to line of sight effects of $\sim$0.075 \textsl{mag}. 
For the Fornax cluster, Blakeslee et~al.~(2009) estimated a depth of 
2.0$^{+0.4}_{-0.6}$ Mpc ($\pm 2\sigma$ of the distance distribution in the 
line-of-sight), equivalent to a dispersion of $\sim$0.05 \textsl{mag}.
Therefore for both clusters, the observed dispersion is
significantly higher than the one expected from the cluster depth only.
 
Given the observational errors and the known depths of the two
clusters, we would like to determine whether there is any intrinsic
dispersion in the value of $\mu$. In order to do that we simulated a
distribution of N galaxies (with N being 89 and 43 for Virgo and
Fornax respectively) with roughly the same intrinsic turnover
magnitude (we included a slight trend in luminosity derived from the
lower panel of Figure \ref{fig:tover_ab}), and we assigned them a
random distance by using a Gaussian depth distribution with
appropriate width (0.075 \textsl{mag} for Virgo and 0.05 \textsl{mag} for Fornax). An
additional random error was added to this distribution based on the
observed uncertainties of our samples. The final distribution of
magnitudes was then used to measure the dispersion of the simulated
sample also by using a 3-sigma clipping algorithm. This procedure was
iterated 10000 times for each sample, delivering a mean expected
dispersion in the value of $\mu$ of 0.22 \textsl{mag} for Virgo, and 0.23 
\textsl{mag} for Fornax. These values are lower than the dispersion
measured in our samples, so we added an additional intrinsic
dispersion term to the simulations until the observed dispersion was
reached. This difference allows for an additional dispersion of 0.21
\textsl{mag} in the case of Virgo and 0.15 \textsl{mag} for the Fornax 
cluster, which can not be accounted by the cluster depth or the observational 
errors alone, and therefore corresponds to an intrinsic dispersion in the
value of $\mu$.

The $z$-band histograms shown in Figure \ref{fig:MTO} are not
symmetric around zero, a higher dispersion can be observed for
positive values of $(\mu_{z}-\bar{\mu}_{z})$. This is consistent with
the fact that the GCLF parameters will always be more precisely
determined for galaxies with larger GC systems and they dominate the
estimation of an error-weighted mean. As we will discuss in
\S\ref{sec:StCand} low luminosity galaxies tend to show fainter
turnover magnitudes and will be therefore located on the positive side
in Figure \ref{fig:MTO}, which combined with the larger uncertainty on
the determination of $\mu_{0}$ in these systems is responsible for the 
larger scatter for the positive values of $(\mu_{z}-\bar{\mu}_{z})$. 
We stress that, as mentioned above, in the simulations done to estimate 
the intrinsic dispersion this slight trend of $\mu$ with $M_{z,gal}$ is taken 
into account.

\section{The Universality of $M_{TO}$} 
\label{sec:StCand}
The use of the GCLF as a distance indicator is based on the assumption
of a universal value of $M_{TO}$, which has indeed been shown to be fairly
constant (within $\pm$0.2 mag for massive galaxies) for a wide range
of galaxy environments. The precision and quantity of our observations
allow to probe for potential dependencies of $M_{TO}$ on factors
such us the luminosity of the parent galaxy, Hubble type, mean color
of the GC system, and environment, that might lurk in the observed
first-order constancy of $M_{TO}$.

Probing for a dependence on Hubble type is important because the usual
procedure is to use the Milky Way and M31 (both spiral galaxies) data
in order to calibrate the GCLF in distant ellipticals.  Our sample
consists exclusively of early-type galaxies, so we cannot study the
effect that the Hubble type might have on the value of
$M_{TO}$. However, we will discuss this later from the point of view
of the metallicity, as the differences in the GCLF as function of the
Hubble type have been attributed to metallicity variations between the
galaxies (Ashman et~al., 1995). We will now address the influence of
these factors on our observed non-universal GCLF.

\subsection{Luminosity}
\label{sec:lum}
The question of whether bright galaxies do have the same $M_{TO}$ as
faint galaxies is particularly interesting to study now that the
correlation between $\sigma$ and $M_{\lowercase{gal}}$ has been
clearly established. Whitmore (1997) has claimed that dwarf
ellipticals have values of $M_{TO}^{V}$ which are roughly 0.3 \textsl{mag}
fainter than bright ellipticals, which was previously also mentioned
by Durrell et~al. (1996). In principle this should not
represent a problem for the use of the GCLF as a distance indicator,
as the method is mostly concentrated on massive galaxies which can be
traced to larger distances. Jord\'an et~al.~(2006, 2007b) have also
noticed that the turnover mass is slightly smaller in dwarf systems
($M_{B}\geq -18$) compared to more massive galaxies (see also Miller 
\& Lotz (2007), showing that this might be partly accounted for by 
the effects of dynamical friction.

\begin{figure}
\begin{center}
\includegraphics[width=8.3cm]{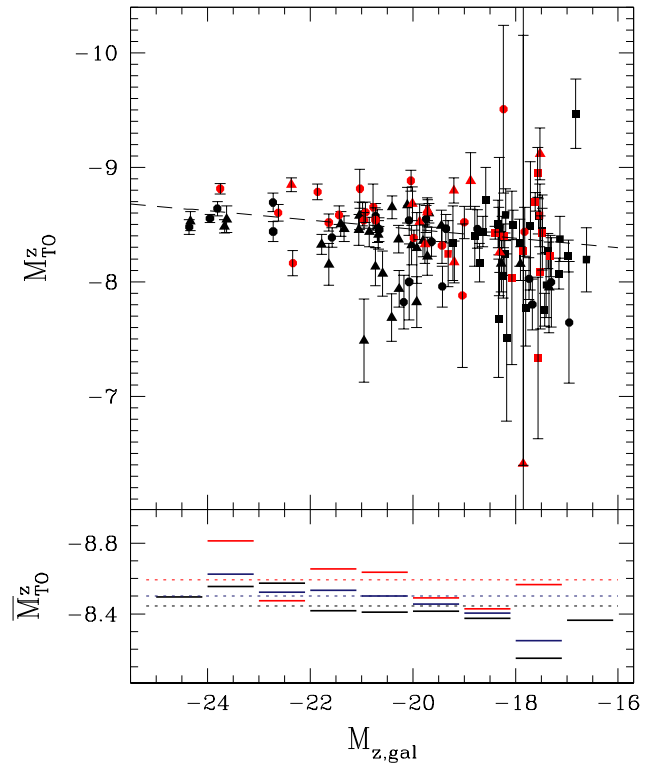}
\caption[GCLF absolute turnover magnitude ($M^{z}_{TO}$)
vs. $M_{B,gal}$] {Top: GCLF absolute $z$-band turnover magnitude
($M^{z}_{TO}$) derived from SBF distances (Blakeslee et~al.~2009) vs. 
the absolute $z$-band magnitude of the parent galaxy ($M_{z,gal}$), 
for all the galaxies in the ACS Virgo (black) and Fornax (red) 
cluster surveys. The error-weighted linear fit corresponds to:
$M^{z}_{TO} = (-7.66 \pm 0.18) + (0.04 \pm 0.01)M_{z,gal}$. The sample 
has been morphologically separated into elliptical (circles), 
lenticular (triangles), and dwarf (squares) galaxies.
Bottom: Weighted mean turnover magnitude ($\bar{\textrm M}$$^{z}_{TO}$) 
calculated in 1\textsl{mag} wide bins (continuous lines) and over the 
whole magnitude range (dotted lines), for the Virgo (black) and Fornax 
(red) sample. The blue lines also show the corresponding weighted mean 
values but using the combined sample.}
\label{fig:tover_ab}
\end{center}
\end{figure}

We investigate a possible dependence of $\mu$ on $M_{z,gal}$ in Figure
\ref{fig:tover_ab}, which is equivalent to Figure \ref{fig:tover_err}
but with the observed turnover magnitudes now transformed to absolute 
turnover magnitudes using SBF distances (Mei et~al.~2007, 
Blakeslee et~al.~2009). The observed values of $\mu$ are relatively 
homogeneous in the range of $M_{z,gal}$ covered by our observations, 
between \mbox{$M_{z,\lowercase{gal}} \sim -24$} and
\mbox{$M_{z,\lowercase{gal}} \sim -17$}, however a tendency for dwarf
galaxies to show slightly less luminous turnover magnitudes seems to
be present. This tendency is characterized by the linear fit:
$M^{z}_{TO} = (-7.66 \pm 0.18) + (0.04 \pm 0.01)M_{z,gal}$.
The interpretation of this trend needs to be considered
carefully because, due to their low luminosity, dwarf galaxies have
smaller GC systems and the uncertainties on the determination of
$\mu_{TO}$ are therefore higher. In order to lessen this problem, in the
lower panel of Figure~\ref{fig:tover_ab} we have plotted the weighted
mean absolute magnitude in intervals of 1 magnitude compared to the
weighted mean absolute magnitude calculated over the whole range of
magnitudes. The lower luminosity bins tend to have mean magnitudes
that are lower than the general mean both in each cluster and in the
combined sample. At the lower luminosity bin (in the range between
\mbox{$-18 < M_{z,\lowercase{gal}} < -17$}) the weighted mean absolute
magnitude is 0.18 \textsl{mag} lower than the general value of -8.51, 
and 0.3\textsl{mag} lower than the most luminous bin (\mbox{-24 $<
M_{z,\lowercase{gal}} <$ -23}). From Figure \ref{fig:tover_ab} we can
confirm then the trend suggested by Whitmore (1997) and reported by
Jord\'an et~al.(2006, 2007b), and we conclude that the luminosity
(i.e. mass) of the parent galaxy has an effect on determining the peak
of the GCLF, with fainter (lower-mass) galaxies having a fainter GCLF
turnover.

Limiting the analysis to only the most massive galaxies in the sample
($M_{z,\lowercase{gal}} <$ -21) we obtain an average turnover magnitude
of $M_z=-8.53$ \textsl{mag} with a dispersion of 0.18
\textsl{mag}. These are the galaxies that could potentially be used as
a distance indicator, and we can see here that they would deliver an
accurate distance modulus estimation within the cosmic scatter of $\pm
0.2$ \textsl{mag}.  There are nonetheless environmental dependencies
that need to be considered before extending these findings to other
systems because, as we can also observe from
Figure~\ref{fig:tover_ab}, the galaxies in the Fornax cluster show
absolute turnover magnitudes that are systematically brighter than the
Virgo sample.  We discus this point further in section
\S\ref{sec:environment}.

\subsection{Color}
\label{sec:color}
One of the most important requirements that a galaxy needs to fulfill
in order to make feasible the use of its GCLF as a distance estimator
is that its GC population must be old. The presence of an
intermediate-age population will modify the GCLF by introducing
clusters that will have brighter magnitudes than the older population.

The GC color distribution of our sample of 89 galaxies in Virgo was
presented by Peng et~al.~(2006), where it was observed that on average
galaxies at all luminosities in the samples (\mbox{$-24 <
M_{z,\lowercase{gal}} < -17$}) appear to have bimodal or asymmetric GC
color distributions. As discussed in Villegas et~al.~(2010, in
preparation) the use of stellar population models allow us to discard
large age differences between red and blue GCs if we assume that the
mass distribution of GCs does not have a dependence on
$[Fe/H]$ inside a given galaxy. With only a few exceptions, the 
population of blue and red GC appear to be coeval within errors 
for most of the galaxies, which lead us to concentrate on the 
problem of different metallicities between them. For giant ellipticals, 
this is also supported by previous
observational studies (Puzia et~al.~1999; Beasley et~al.~2000,
Jord\'an et~al.~2002), although there are examples of massive galaxies
that appear to have formed GCs recently triggered by mergers (e.g.~
NGC~1316, Goudfrooij et~al.~2001). 
 
With the goal of obtaining an improved calibration for the value of
$M_{TO}$, Ashman et~al.~(1995) studied the effects of metallicity on
the GCLF showing that changes in the mean metallicity of the cluster
sample produce a shift on $M_{TO}$, provided the mass distribution
does not depend on [Fe/H]. According to Bruzual \& Charlot (2003) 
models, the expected change in $z$-band turnover magnitude, $M^{z}_{TO}$, 
over the range of GC mean metallicity is $<$ 0.02 \textsl{mag},  
which is utterly negligible considering the observational errors.

\begin{figure}
\centering
\includegraphics[scale=0.8]{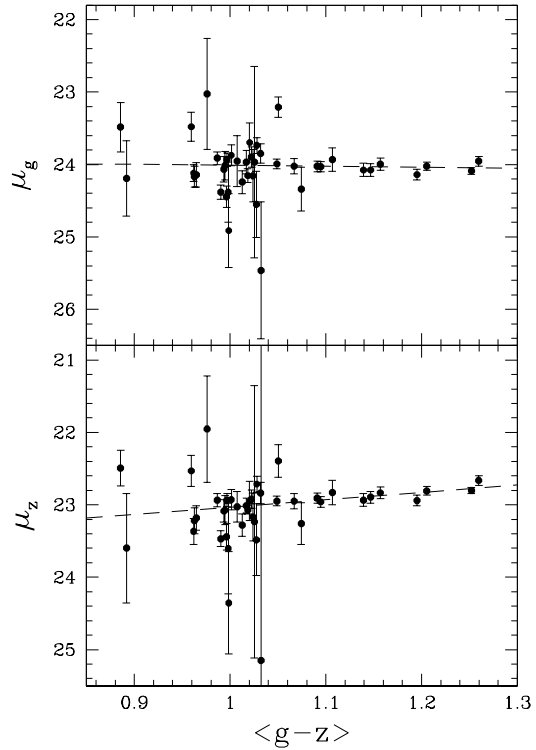}
\caption{Turnover magnitude vs. mean GC color relationship for
our Fornax cluster data in $g$ (top) and $z$-bands (bottom).
The lines show the error weighted fit to the data and have a slopes of
$0.13 \pm 0.27$ in the $g$-band and $-1.01 \pm 0.22$ in the $z$-band.}
\label{fig:col_john}
\end{figure}

From a different point of view Figure \ref{fig:col_john} shows the 
correlation between turnover magnitude $\mu$ and mean GC color 
$\langle g-z \rangle$, in both $g$ (top) and $z$-band (bottom) for all 
the galaxies in the Fornax sample. From this plot it can be observed 
that on average $\mu_g$ remains constant as a function of 
$\langle g-z \rangle$, but $\mu_z$ tends to be brighter for redder GC 
systems. The interpretation of this plot presents a degeneracy between age 
and mass. If we assume that the Fornax galaxies, and by extension their 
GC systems, are all basically coeval, this trend can be explained by the 
fact that the $z$-band turnover better reflects mass (as it is only loosely 
dependent on metallicity), and therefore this is an indication that galaxies 
with lower masses (as accounted by the mean metallicity of its GC system)
might have less-massive turnover values, which translates into fainter 
$\mu_z$. In the $g$-band, and as a consequence also in the nearby V-band, 
this effect is canceled by the fact that the mass-to-light ratio gets 
lower for GCs in lower-mass, lower-metallicity galaxies. 
Therefore the historically ``constant turnover magnitude'' of the V-band 
GCLF might just be a consequence of the incidental cancellation of these 
two factors at this wavelength.

\subsection{Environment}
\label{sec:environment}

Even if we assume that the GC populations of galaxies of all
morphological types are formed with the same initial mass function
irrespective of available gas mass and metallicity, there is still the
environmental factor to play against the existence of a universal
GCLF. The particular media in which the clusters are formed might
affect their evolution, shaping an environment-dependent GCLF.

Based on data from groups and clusters of galaxies Blakeslee \& Tonry
(1996) found evidence that $M_{TO}$ becomes fainter as the local
density of galaxies increases. They used the velocity dispersion of
groups of galaxies in the local universe as a density indicator in
order to compare the values of $M_{TO}$ in different environments. 
Our data support the evidence presented by Blakeslee \& Tonry (1996) 
in the sense they have also found a
relative distance modulus that is too small compared to the SBF
measurements. The trend of $M_{TO}$ changing as a function of
environment (as accounted by velocity dispersion) is also followed by
our data. However, it is important to mention that in spite of
its lower velocity dispersion the Fornax cluster is denser than Virgo
(Ferguson 1989b), and therefore the observed tendency seems to be more 
related to the total mass of the cluster than to local density.

Also, as discussed in \S\ref{sec:dist_in} the observed discrepancy on
the estimation of the relative Virgo-Fornax distance could be
interpreted as a difference in mean-age between the stellar
populations of these two clusters of galaxies, with an age difference
of 3 Gyr being enough to explain this discrepancy.  The combined use
of high-resolution cosmological simulations and semi-analytic
techniques (De Lucia et~al.~2006) has shown that the faster evolution
of protocluster regions produces star formation histories that peak at
higher redshift for early-type galaxies hosted by more massive halos.
This effect would therefore produce stellar populations in the Virgo
clusters that are on average older when compared to stellar population
belonging to a less massive galaxy cluster like Fornax. Mass estimates
for Virgo vary substantially (e.g., $(0.15-1.5)\times 10^{15}
\mathcal{M}_{\odot}$; B\"ohringer et al.~1994; Schindler et~al.~1999;
McLaughlin 1999; Tonry et~al.~2000; Fouqu\'e et~al.~2001), but it is
clear that its total mass is nearly an order of magnitude higher than
the mass of Fornax ($\sim 7 \times 10^{13} \mathcal{M}_{\odot}$,
Drinkwater et~al.~2001).  The results presented by De Lucia
et~al.~(2006) predict the expected mean-age difference for clusters of
these masses to be $\sim$0.5 Gyr, a value that is too low to explain
the observed difference in turnover magnitudes, but that is also
dependant on the input parameters of the simulations.

\section{Summary and Conclusions} 
\label{sec:CONC}
We used ACS/HST data in order to study the GCLF of 89 early-type
galaxies in the Virgo cluster and 43 galaxies in the Fornax cluster,
which constitute the most homogeneous set of data used to date for
this purpose. The GCLF of these galaxies was fitted by using a maximum
likelihood approach to model the intrinsic Gaussian distribution after
accounting for contamination and completeness effects. From the
derived values of the turnover magnitude and the dispersion of the
Gaussian fits we conclude that:

\begin{enumerate}

\item The analysis of 43 early-type galaxies belonging to the Fornax
  cluster shows that the dispersion of the GCLF decreases as the
  luminosity of the host galaxy decreases, confirming our previous
  results obtained with Virgo galaxies (Jord\'an et~al. 2006, 2007b). 
  
\item By using the GCLF turnover magnitude as a distance indicator on
  our homogeneous data set we derive a relative distance modulus between
  the Virgo and the Fornax clusters of $\Delta(m-M)_{\rm GCLF}=0.20\pm 0.04$ 
  \textsl{mag}, which is lower than the one derived using SBF measurements on 
  the same data, $\Delta (m-M)_{\rm SBF}=0.42\pm0.03$ \textsl{mag}.

\item Setting the relative Virgo-Fornax distance as that given by SBF
  implies a difference in the value of $\langle \mu_{TO} \rangle$ in
  the two closest clusters of galaxies, suggesting that this quantity
  is influenced by the environment in which a GC system is formed
  and evolves. These results support a previous study by Blakeslee \&
  Tonry (1996), who found a correlation between GCLF turnover magnitude
  and velocity dispersion of the host cluster, in the sense that galaxy
  clusters with higher velocity dispersions (higher masses) host
  galaxies with fainter turnovers in their GC systems.

\item The discrepancy in the absolute magnitude of the GCLF turnovers
  in Virgo and Fornax can be accounted for if GC systems in the Fornax
  clusters were on average $\sim$3 Gyrs younger than those in Virgo
  (thus making them brighter). Recent results from high-resolution
  numerical simulations (e.g.~Springel et~al.~2005, De Lucia et~al.~2006) 
  suggest that stellar populations of Virgo-like galaxy
  clusters (high mass and high velocity dispersion) were formed mostly
  at higher redshift compared to less massive and lower-dispersion
  clusters like Fornax. This trend could therefore be at least partially
  responsible for the observed discrepancy in the absolute GCLF turnover 
  magnitudes between both clusters.

\item We have measured a total dispersion on the value of the turnover
   magnitude of 0.31 and 0.28 \textsl{mag} for Virgo and Fornax
   respectively. We show using simulations that these values can be
   only partially accounted by the dispersion produced by cluster
   depth and observational uncertainties. The additional dispersion
   can be modeled by an intrinsic dispersion on the value of $\mu_{0}$
   of 0.21 \textsl{mag} for the Virgo cluster and 0.15 \textsl{mag} for 
   Fornax.

\item The measured GCLF turnover is found to be systematically fainter
   for low luminosity galaxies, showing a $\sim$0.3 \textsl{mag} decrease 
   on dwarf systems, although we suffer from large uncertainties in that
   galaxy luminosity regime. The luminosity (i.e.~$\sim$ mass) of the
   parent galaxy seems to play an important role on shaping the final
   form of the luminosity distribution. This might be at least partly
   accounted for by the effects of dynamical friction if all other
   processes that contribute on shaping the mass function (two-body
   relaxation, tidal shocks, etc.)  were to lead to a roughly constant
   M$_{TO}$ (Jord\'an et~al. 2007b).

\item Overall we find that GCLF parameters vary continuously and 
  systematically as a function of galaxy luminosity (i.e.~mass). The 
  correlations we present here show no evidence for a dichotomy between 
  giant and dwarf early-type galaxies at $M_z \sim -19.5$ 
  ($M_B \sim -18$) in terms of their GC systems. This is 
  consistent with results presented in several recent studies (e.g.~Graham 
  \& Guzm\'an, 2003; Gavazzi et~al.~2005, C\^ot\'e et~al.,~2006), 
  and is at odds with earlier claims by Kormendy (1985).

\end{enumerate}  

\acknowledgments
\vspace*{2cm} Support for programs GO-9401 and GO-10217 was provided
through a grant from the Space Telescope Science Institute, which is
operated by the Association of Universities for Research in Astronomy,
Inc., under NASA contract NAS5-26555. A.J.~and L.I.~acknowledge support
from the Chilean Center of Excellence in Astrophysics and Associated
Technologies (PFB 06) and from the Chilean Center for Astrophysics
FONDAP 15010003. Additional support for A.J.~is
provided by MIDEPLAN's Programa Inicativa Cient\'{i}fica Milenio
through grant P07-021-F. This research has made use of the NASA/IPAC
Extragalactic Database (NED) which is operated by the Jet Propulsion
Laboratory, California Institute of Technology, under contract with
the National Aeronautics and Space Administration.

\clearpage
\LongTables
\begin{deluxetable*}{lrrccccrrc}
\tablecolumns{8}
\tablewidth{0pc}
\linespread{2.5}
\tablecaption{Gaussian GCLF parameters for all ACSVCS galaxies.}
\tablehead{
\colhead{ID} & \colhead{$B_{gal}$} &  \colhead{$\mu_g$} & \colhead{$\sigma_{g}$} & \colhead{$\mu_{z}$} & \colhead{$\sigma_{z}$} 
& \colhead{$\hat \beta$} & \colhead{$N$} \\ 
\colhead{(1)} & \colhead{(2)} & \colhead{(3)} & \colhead{(4)} & \colhead{(5)} & \colhead{(6)}& \colhead{(7)} & \colhead{(8)} }\\ 
\startdata
VCC 1226 &  9.31 &  23.947 $\pm$ 0.066 & 1.340 $\pm$ 0.050 & 22.670 $\pm$ 0.063 & 1.304 $\pm$ 0.048 & 0.023 & 765  \\
VCC 1316 &  9.58 &  23.872 $\pm$ 0.039 & 1.283 $\pm$ 0.030 & 22.591 $\pm$ 0.036 & 1.223 $\pm$ 0.028 & 0.014 & 1745 \\
VCC 1978 &  9.81 &  23.893 $\pm$ 0.059 & 1.296 $\pm$ 0.046 & 22.636 $\pm$ 0.059 & 1.293 $\pm$ 0.046 & 0.022 & 807  \\
VCC 881  & 10.06 &  23.887 $\pm$ 0.087 & 1.280 $\pm$ 0.068 & 22.775 $\pm$ 0.083 & 1.240 $\pm$ 0.066 & 0.034 & 367  \\
VCC 798  & 10.09 &  23.889 $\pm$ 0.115 & 1.194 $\pm$ 0.078 & 22.760 $\pm$ 0.116 & 1.157 $\pm$ 0.080 & 0.012 & 370  \\
VCC 763  & 10.26 &  23.874 $\pm$ 0.063 & 1.155 $\pm$ 0.050 & 22.759 $\pm$ 0.063 & 1.145 $\pm$ 0.049 & 0.035 & 506  \\
VCC 731  & 10.51 &  24.343 $\pm$ 0.055 & 1.201 $\pm$ 0.043 & 23.166 $\pm$ 0.055 & 1.198 $\pm$ 0.043 & 0.021 & 907  \\
VCC 1535 & 10.61 &  23.664 $\pm$ 0.087 & 1.107 $\pm$ 0.068 & 22.503 $\pm$ 0.086 & 1.091 $\pm$ 0.067 & 0.042 & 244  \\
VCC 1903 & 10.76 &  23.405 $\pm$ 0.078 & 1.175 $\pm$ 0.063 & 22.214 $\pm$ 0.081 & 1.198 $\pm$ 0.065 & 0.046 & 308  \\
VCC 1632 & 10.78 &  23.860 $\pm$ 0.089 & 1.400 $\pm$ 0.069 & 22.643 $\pm$ 0.086 & 1.374 $\pm$ 0.067 & 0.038 & 456  \\
VCC 1231 & 11.10 &  23.710 $\pm$ 0.084 & 1.112 $\pm$ 0.065 & 22.571 $\pm$ 0.084 & 1.105 $\pm$ 0.065 & 0.058 & 254  \\
VCC 2095 & 11.18 &  24.616 $\pm$ 0.321 & 1.669 $\pm$ 0.203 & 23.638 $\pm$ 0.363 & 1.693 $\pm$ 0.221 & 0.076 & 134  \\
VCC 1154 & 11.37 &  23.887 $\pm$ 0.085 & 0.993 $\pm$ 0.066 & 22.763 $\pm$ 0.087 & 0.990 $\pm$ 0.067 & 0.065 & 192  \\
VCC 1062 & 11.40 &  23.638 $\pm$ 0.114 & 1.208 $\pm$ 0.089 & 22.495 $\pm$ 0.112 & 1.187 $\pm$ 0.088 & 0.066 & 179  \\
VCC 2092 & 11.51 &  24.030 $\pm$ 0.172 & 1.127 $\pm$ 0.133 & 22.923 $\pm$ 0.184 & 1.175 $\pm$ 0.139 & 0.114 & 92   \\
VCC 369  & 11.80 &  23.609 $\pm$ 0.102 & 1.101 $\pm$ 0.079 & 22.414 $\pm$ 0.099 & 1.062 $\pm$ 0.079 & 0.068 & 179  \\
VCC 759  & 11.80 &  23.803 $\pm$ 0.110 & 1.130 $\pm$ 0.089 & 22.687 $\pm$ 0.107 & 1.100 $\pm$ 0.086 & 0.067 & 172  \\
VCC 1692 & 11.82 &  23.791 $\pm$ 0.123 & 1.051 $\pm$ 0.095 & 22.747 $\pm$ 0.135 & 1.099 $\pm$ 0.104 & 0.096 & 136  \\
VCC 1030 & 11.84 &  23.711 $\pm$ 0.090 & 0.980 $\pm$ 0.070 & 22.595 $\pm$ 0.092 & 1.013 $\pm$ 0.071 & 0.072 & 176  \\
VCC 2000 & 11.94 &  23.511 $\pm$ 0.107 & 1.201 $\pm$ 0.082 & 22.471 $\pm$ 0.104 & 1.163 $\pm$ 0.080 & 0.071 & 197  \\
VCC 685  & 11.99 &  23.639 $\pm$ 0.121 & 1.236 $\pm$ 0.095 & 22.555 $\pm$ 0.120 & 1.210 $\pm$ 0.098 & 0.085 & 167  \\
VCC 1664 & 12.02 &  23.665 $\pm$ 0.109 & 1.059 $\pm$ 0.085 & 22.472 $\pm$ 0.103 & 1.009 $\pm$ 0.083 & 0.092 & 146  \\
VCC 654  & 12.03 &  23.991 $\pm$ 0.183 & 0.926 $\pm$ 0.135 & 23.056 $\pm$ 0.198 & 0.940 $\pm$ 0.152 & 0.194 & 48   \\
VCC 944  & 12.08 &  23.708 $\pm$ 0.121 & 0.872 $\pm$ 0.093 & 22.651 $\pm$ 0.124 & 0.864 $\pm$ 0.097 & 0.132 & 91   \\
VCC 1938 & 12.11 &  23.766 $\pm$ 0.133 & 1.076 $\pm$ 0.110 & 22.792 $\pm$ 0.128 & 1.009 $\pm$ 0.120 & 0.114 & 101  \\
VCC 1279 & 12.15 &  23.645 $\pm$ 0.105 & 1.026 $\pm$ 0.079 & 22.621 $\pm$ 0.111 & 1.048 $\pm$ 0.085 & 0.097 & 138  \\
VCC 1720 & 12.29 &  23.670 $\pm$ 0.127 & 0.797 $\pm$ 0.102 & 22.613 $\pm$ 0.143 & 0.870 $\pm$ 0.115 & 0.141 & 71   \\
VCC 355  & 12.41 &  24.504 $\pm$ 0.279 & 1.208 $\pm$ 0.207 & 23.316 $\pm$ 0.206 & 1.027 $\pm$ 0.158 & 0.167 & 62   \\
VCC 1619 & 12.50 &  24.261 $\pm$ 0.219 & 1.074 $\pm$ 0.161 & 23.166 $\pm$ 0.234 & 1.082 $\pm$ 0.171 & 0.165 & 66   \\
VCC 1883 & 12.57 &  24.125 $\pm$ 0.187 & 1.135 $\pm$ 0.148 & 22.996 $\pm$ 0.166 & 1.064 $\pm$ 0.136 & 0.124 & 83   \\
VCC 1242 & 12.60 &  23.731 $\pm$ 0.113 & 0.927 $\pm$ 0.088 & 22.636 $\pm$ 0.120 & 0.983 $\pm$ 0.093 & 0.105 & 116  \\
VCC 784  & 12.67 &  24.269 $\pm$ 0.161 & 0.865 $\pm$ 0.123 & 23.102 $\pm$ 0.159 & 0.806 $\pm$ 0.131 & 0.179 & 64   \\
VCC 1537 & 12.70 &  23.662 $\pm$ 0.240 & 0.977 $\pm$ 0.183 & 22.750 $\pm$ 0.309 & 1.124 $\pm$ 0.232 & 0.256 & 45   \\
VCC 778  & 12.72 &  24.073 $\pm$ 0.178 & 1.052 $\pm$ 0.139 & 22.972 $\pm$ 0.172 & 1.009 $\pm$ 0.134 & 0.163 & 74   \\
VCC 1321 & 12.84 &  24.160 $\pm$ 0.225 & 0.926 $\pm$ 0.168 & 23.153 $\pm$ 0.222 & 0.919 $\pm$ 0.166 & 0.198 & 50   \\
VCC 828  & 12.84 &  23.804 $\pm$ 0.157 & 1.045 $\pm$ 0.142 & 22.787 $\pm$ 0.131 & 0.895 $\pm$ 0.113 & 0.143 & 80   \\
VCC 1250 & 12.91 &  23.583 $\pm$ 0.145 & 0.815 $\pm$ 0.111 & 22.609 $\pm$ 0.154 & 0.831 $\pm$ 0.118 & 0.200 & 54   \\
VCC 1630 & 12.91 &  24.124 $\pm$ 0.326 & 1.283 $\pm$ 0.232 & 23.104 $\pm$ 0.331 & 1.304 $\pm$ 0.230 & 0.217 & 57   \\
VCC 1146 & 12.93 &  23.939 $\pm$ 0.141 & 0.970 $\pm$ 0.186 & 22.749 $\pm$ 0.127 & 0.890 $\pm$ 0.124 & 0.148 & 82   \\
VCC 1025 & 13.06 &  24.251 $\pm$ 0.112 & 0.847 $\pm$ 0.097 & 23.335 $\pm$ 0.136 & 0.938 $\pm$ 0.110 & 0.143 & 104  \\
VCC 1303 & 13.10 &  23.681 $\pm$ 0.140 & 0.821 $\pm$ 0.106 & 22.793 $\pm$ 0.139 & 0.805 $\pm$ 0.108 & 0.176 & 61   \\
VCC 1913 & 13.22 &  23.688 $\pm$ 0.113 & 0.724 $\pm$ 0.103 & 22.675 $\pm$ 0.117 & 0.738 $\pm$ 0.102 & 0.181 & 65   \\
VCC 1327~* & 13.26 &  23.688 $\pm$ 0.121 & 1.262 $\pm$ 0.093 & 22.626 $\pm$ 0.115 & 1.212 $\pm$ 0.088 & 0.081 & 173 \\ 
VCC 1125 & 13.30 &  23.667 $\pm$ 0.127 & 0.781 $\pm$ 0.109 & 22.645 $\pm$ 0.136 & 0.791 $\pm$ 0.109 & 0.179 & 62   \\
VCC 1475 & 13.36 &  24.073 $\pm$ 0.141 & 0.990 $\pm$ 0.107 & 23.199 $\pm$ 0.178 & 1.101 $\pm$ 0.133 & 0.137 & 86   \\
VCC 1178 & 13.37 &  23.609 $\pm$ 0.134 & 0.997 $\pm$ 0.102 & 22.562 $\pm$ 0.123 & 0.949 $\pm$ 0.090 & 0.124 & 90   \\
VCC 1283 & 13.45 &  24.049 $\pm$ 0.152 & 0.894 $\pm$ 0.120 & 23.023 $\pm$ 0.167 & 0.932 $\pm$ 0.129 & 0.170 & 66   \\
VCC 1261 & 13.56 &  23.962 $\pm$ 0.275 & 1.133 $\pm$ 0.208 & 23.004 $\pm$ 0.327 & 1.243 $\pm$ 0.238 & 0.217 & 46   \\
VCC 698  & 13.60 &  23.793 $\pm$ 0.090 & 0.843 $\pm$ 0.066 & 22.777 $\pm$ 0.085 & 0.810 $\pm$ 0.062 & 0.105 & 119  \\
VCC 1422 & 13.64 &  23.625 $\pm$ 0.169 & 0.656 $\pm$ 0.130 & 22.521 $\pm$ 0.168 & 0.651 $\pm$ 0.127 & 0.258 & 37   \\
VCC 2048 & 13.81 &  23.450 $\pm$ 0.324 & 0.969 $\pm$ 0.217 & 22.420 $\pm$ 0.282 & 0.881 $\pm$ 0.194 & 0.420 & 22   \\
VCC 1871 & 13.86 &  23.520 $\pm$ 0.608 & 1.181 $\pm$ 0.455 & 22.512 $\pm$ 0.604 & 1.154 $\pm$ 0.480 & 0.516 & 18   \\
VCC 9    & 13.93 &  23.940 $\pm$ 0.391 & 1.086 $\pm$ 0.305 & 22.830 $\pm$ 0.260 & 0.894 $\pm$ 0.196 & 0.246 & 34   \\
VCC 575  & 14.14 &  24.847 $\pm$ 0.271 & 0.665 $\pm$ 0.281 & 23.833 $\pm$ 0.130 & 0.333 $\pm$ 0.184 & 0.386 & 27   \\
VCC 1910 & 14.17 &  23.758 $\pm$ 0.208 & 1.175 $\pm$ 0.161 & 22.630 $\pm$ 0.209 & 1.135 $\pm$ 0.176 & 0.180 & 60   \\
VCC 1049 & 14.20 &  24.052 $\pm$ 0.257 & 0.550 $\pm$ 0.197 & 23.106 $\pm$ 0.396 & 0.634 $\pm$ 0.268 & 0.487 & 18   \\
VCC 856  & 14.25 &  23.792 $\pm$ 0.185 & 0.887 $\pm$ 0.156 & 22.768 $\pm$ 0.164 & 0.862 $\pm$ 0.127 & 0.211 & 50   \\
VCC 140  & 14.30 &  23.992 $\pm$ 0.245 & 0.790 $\pm$ 0.197 & 22.979 $\pm$ 0.249 & 0.822 $\pm$ 0.182 & 0.329 & 29   \\
VCC 1355 & 14.31 &  24.554 $\pm$ 0.776 & 1.273 $\pm$ 0.541 & 23.682 $\pm$ 0.732 & 1.161 $\pm$ 0.530 & 0.471 & 20   \\
VCC 1087 & 14.31 &  23.732 $\pm$ 0.134 & 0.926 $\pm$ 0.101 & 22.713 $\pm$ 0.133 & 0.898 $\pm$ 0.112 & 0.162 & 68   \\
VCC 1297~* & 14.33 &  23.403 $\pm$ 0.109 & 1.141 $\pm$ 0.082 & 22.299 $\pm$ 0.105 & 1.084 $\pm$ 0.080 & 0.092 & 152 \\ 
VCC 1861 & 14.37 &  23.608 $\pm$ 0.222 & 1.015 $\pm$ 0.185 & 22.572 $\pm$ 0.206 & 0.937 $\pm$ 0.164 & 0.234 & 49   \\
VCC 543  & 14.39 &  23.854 $\pm$ 0.196 & 0.692 $\pm$ 0.139 & 22.792 $\pm$ 0.184 & 0.635 $\pm$ 0.127 & 0.330 & 28   \\
VCC 1431 & 14.51 &  24.092 $\pm$ 0.171 & 1.054 $\pm$ 0.128 & 23.054 $\pm$ 0.188 & 1.082 $\pm$ 0.140 & 0.158 & 71   \\
VCC 1528 & 14.51 &  23.550 $\pm$ 0.137 & 0.720 $\pm$ 0.105 & 22.609 $\pm$ 0.129 & 0.697 $\pm$ 0.097 & 0.222 & 49   \\
VCC 1695 & 14.53 &  24.416 $\pm$ 0.401 & 0.962 $\pm$ 0.289 & 23.480 $\pm$ 0.517 & 1.103 $\pm$ 0.357 & 0.380 & 22   \\
VCC 1833 & 14.54 &  24.091 $\pm$ 0.223 & 0.695 $\pm$ 0.159 & 22.954 $\pm$ 0.147 & 0.500 $\pm$ 0.110 & 0.332 & 28   \\
VCC 437  & 14.54 &  23.933 $\pm$ 0.162 & 0.783 $\pm$ 0.134 & 23.056 $\pm$ 0.167 & 0.845 $\pm$ 0.131 & 0.229 & 50   \\
VCC 2019 & 14.55 &  23.551 $\pm$ 0.220 & 0.873 $\pm$ 0.200 & 22.619 $\pm$ 0.225 & 0.860 $\pm$ 0.193 & 0.303 & 34   \\
VCC 200  & 14.69 &  24.459 $\pm$ 0.221 & 0.680 $\pm$ 0.144 & 23.582 $\pm$ 0.331 & 0.834 $\pm$ 0.221 & 0.381 & 25   \\
VCC 571  & 14.74 &  24.392 $\pm$ 0.543 & 0.951 $\pm$ 0.346 & 24.249 $\pm$ 1.542 & 1.421 $\pm$ 0.810 & 0.478 & 17   \\
VCC 21   & 14.75 &  24.073 $\pm$ 0.636 & 1.438 $\pm$ 0.418 & 22.963 $\pm$ 0.559 & 1.276 $\pm$ 0.387 & 0.351 & 26   \\
VCC 1488 & 14.76 &  24.146 $\pm$ 0.303 & 0.580 $\pm$ 0.208 & 23.088 $\pm$ 0.390 & 0.553 $\pm$ 0.262 & 0.471 & 19   \\
VCC 1499 & 14.94 &  24.562 $\pm$ 0.601 & 1.418 $\pm$ 0.377 & 23.489 $\pm$ 0.530 & 1.295 $\pm$ 0.341 & 0.272 & 35   \\
VCC 1545 & 14.96 &  24.099 $\pm$ 0.164 & 0.910 $\pm$ 0.128 & 23.159 $\pm$ 0.183 & 0.930 $\pm$ 0.145 & 0.189 & 63   \\
VCC 1192~* & 15.04 &  23.781 $\pm$ 0.086 & 1.073 $\pm$ 0.066 & 22.663 $\pm$ 0.085 & 1.052 $\pm$ 0.064 & 0.064 & 213\\
VCC 1075 & 15.08 &  23.514 $\pm$ 0.169 & 0.554 $\pm$ 0.119 & 22.522 $\pm$ 0.155 & 0.515 $\pm$ 0.115 & 0.378 & 26   \\
VCC 1440 & 15.20 &  24.267 $\pm$ 0.237 & 0.895 $\pm$ 0.176 & 23.270 $\pm$ 0.221 & 0.824 $\pm$ 0.162 & 0.259 & 38   \\
VCC 230  & 15.20 &  23.941 $\pm$ 0.134 & 0.541 $\pm$ 0.106 & 23.078 $\pm$ 0.139 & 0.578 $\pm$ 0.105 & 0.274 & 38   \\
VCC 2050 & 15.20 &  23.900 $\pm$ 0.118 & 0.281 $\pm$ 0.089 & 22.963 $\pm$ 0.135 & 0.304 $\pm$ 0.106 & 0.459 & 20   \\
VCC 751  & 15.30 &  23.525 $\pm$ 0.191 & 0.504 $\pm$ 0.130 & 22.699 $\pm$ 0.206 & 0.509 $\pm$ 0.130 & 0.495 & 17   \\
VCC 1828 & 15.33 &  23.807 $\pm$ 0.210 & 0.702 $\pm$ 0.183 & 22.757 $\pm$ 0.198 & 0.664 $\pm$ 0.148 & 0.355 & 27   \\
VCC 1407 & 15.49 &  24.397 $\pm$ 0.123 & 0.665 $\pm$ 0.094 & 23.420 $\pm$ 0.144 & 0.745 $\pm$ 0.111 & 0.186 & 60   \\
VCC 1886 & 15.49 &  23.034 $\pm$ 0.715 & 0.971 $\pm$ 0.463 & 21.565 $\pm$ 0.304 & 0.463 $\pm$ 0.215 & 0.622 & 14   \\
VCC 1199~* & 15.50 &  23.833 $\pm$ 0.094 & 1.166 $\pm$ 0.074 & 22.682 $\pm$ 0.089 & 1.125 $\pm$ 0.070 & 0.060 & 228 \\ 
VCC 1539 & 15.68 &  23.813 $\pm$ 0.182 & 0.831 $\pm$ 0.168 & 22.820 $\pm$ 0.199 & 0.901 $\pm$ 0.147 & 0.275 & 43   \\
VCC 1185 & 15.68 &  23.843 $\pm$ 0.172 & 0.693 $\pm$ 0.116 & 22.910 $\pm$ 0.155 & 0.639 $\pm$ 0.105 & 0.292 & 33   \\
VCC 1489 & 15.89 &  23.977 $\pm$ 0.150 & 0.378 $\pm$ 0.129 & 23.157 $\pm$ 0.279 & 0.484 $\pm$ 0.469 & 0.417 & 22   \\
VCC 1661 & 15.97 &  24.040 $\pm$ 0.281 & 0.630 $\pm$ 0.273 & 23.058 $\pm$ 0.285 & 0.614 $\pm$ 0.215 & 0.477 & 19   \\
\enddata                                                                                                              
\tablenotetext{}{Notes -- (1) Galaxy VCC number. (2) Galaxy $B$-band magnitude.  (3) and (4) Maximum likelihood 
estimates of the Gaussian mean $\mu$ and dispersion $\sigma$ of the $g$-band GCLF. (5) and (6) Same as cols.~(3) 
and (4), but for the $z$ band. (7) Fraction of the sample that is expected to be contamination. (8) Total number 
N of all objects (including contaminants and uncorrected for incompleteness) with $p_{GC} \geq 0.5$. (*) 
These galaxies were excluded from the analysis because of their close proximity to massive elliptical galaxies.} 
\label{tab:acsvcs_gclf_par}                                                                                           
\end{deluxetable*}    

\clearpage
\LongTables
\begin{deluxetable*}{lrrccccrrc}
\tablecolumns{8}
\tablewidth{0pc}
\linespread{2.5}
\tablecaption{Gaussian GCLF parameters for all ACSFCS galaxies.}
\tablehead{
\colhead{ID} & \colhead{$B_{gal}$}  & \colhead{$\mu_g$} & \colhead{$\sigma_{g}$} & \colhead{$\mu_{z}$} & \colhead{$\sigma_{z}$} 
& \colhead{$\hat \beta$} & \colhead{$N$} \\
\colhead{(1)} & \colhead{(2)} & \colhead{(3)} & \colhead{(4)} & \colhead{(5)} & \colhead{(6)}& \colhead{(7)} & \colhead{(8)} }\\
\startdata
FCC 21  &  9.4 &  26.350 $\pm$ 1.234 & 2.178 $\pm$ 0.059 & 25.150 $\pm$ 0.668 & 2.189 $\pm$ 0.060 & 0.011 &  647 \\ 
FCC 213 & 10.6 &  24.090 $\pm$ 0.048 & 1.231 $\pm$ 0.038 & 22.802 $\pm$ 0.044 & 1.198 $\pm$ 0.035 & 0.015 & 1074 \\ 
FCC 219 & 10.9 &  24.140 $\pm$ 0.072 & 1.110 $\pm$ 0.058 & 22.940 $\pm$ 0.072 & 1.112 $\pm$ 0.058 & 0.039 &  380 \\ 
NGC 1340& 11.2 &  24.384 $\pm$ 0.098 & 1.124 $\pm$ 0.074 & 23.468 $\pm$ 0.111 & 1.180 $\pm$ 0.082 & 0.039 &  280 \\ 
FCC 167 & 11.3 &  24.023 $\pm$ 0.059 & 1.022 $\pm$ 0.046 & 22.808 $\pm$ 0.060 & 1.044 $\pm$ 0.047 & 0.026 &  424 \\ 
FCC 276 & 11.8 &  24.032 $\pm$ 0.070 & 1.102 $\pm$ 0.063 & 22.960 $\pm$ 0.076 & 1.166 $\pm$ 0.061 & 0.040 &  361 \\ 
FCC 147 & 11.9 &  24.077 $\pm$ 0.085 & 1.197 $\pm$ 0.067 & 22.894 $\pm$ 0.081 & 1.156 $\pm$ 0.064 & 0.047 &  320 \\ 
IC 2006 & 12.2 &  24.076 $\pm$ 0.092 & 0.886 $\pm$ 0.070 & 22.935 $\pm$ 0.089 & 0.868 $\pm$ 0.067 & 0.085 &  132 \\ 
FCC 83  & 12.3 &  24.026 $\pm$ 0.076 & 1.040 $\pm$ 0.058 & 22.906 $\pm$ 0.070 & 0.988 $\pm$ 0.054 & 0.044 &  274 \\ 
FCC 184 & 12.3 &  23.956 $\pm$ 0.067 & 1.029 $\pm$ 0.054 & 22.664 $\pm$ 0.067 & 1.030 $\pm$ 0.054 & 0.042 &  306 \\ 
FCC 63  & 12.7 &  24.023 $\pm$ 0.106 & 1.236 $\pm$ 0.084 & 22.951 $\pm$ 0.108 & 1.233 $\pm$ 0.086 & 0.058 &  231 \\ 
FCC 193 & 12.8 &  23.934 $\pm$ 0.161 & 0.822 $\pm$ 0.123 & 22.830 $\pm$ 0.172 & 0.899 $\pm$ 0.130 & 0.176 &   48 \\ 
FCC 153 & 13.0 &  24.066 $\pm$ 0.175 & 0.947 $\pm$ 0.135 & 23.086 $\pm$ 0.155 & 0.857 $\pm$ 0.119 & 0.161 &   60 \\ 
FCC 170 & 13.0 &  24.016 $\pm$ 0.196 & 1.182 $\pm$ 0.167 & 23.073 $\pm$ 0.202 & 1.195 $\pm$ 0.171 & 0.137 &   71 \\ 
FCC 177 & 13.2 &  23.897 $\pm$ 0.139 & 0.928 $\pm$ 0.108 & 22.923 $\pm$ 0.125 & 0.859 $\pm$ 0.095 & 0.129 &   70 \\ 
FCC 47  & 13.3 &  23.993 $\pm$ 0.068 & 0.988 $\pm$ 0.053 & 22.948 $\pm$ 0.068 & 0.984 $\pm$ 0.054 & 0.044 &  276 \\ 
FCC 43  & 13.5 &  24.342 $\pm$ 0.304 & 1.088 $\pm$ 0.238 & 23.261 $\pm$ 0.289 & 1.099 $\pm$ 0.220 & 0.208 &   37 \\ 
FCC 190 & 13.5 &  23.934 $\pm$ 0.090 & 0.932 $\pm$ 0.072 & 22.937 $\pm$ 0.091 & 0.940 $\pm$ 0.073 & 0.071 &  156 \\ 
FCC 310 & 13.5 &  24.144 $\pm$ 0.167 & 0.743 $\pm$ 0.122 & 23.184 $\pm$ 0.169 & 0.736 $\pm$ 0.123 & 0.229 &   39 \\ 
FCC 148 & 13.6 &  23.851 $\pm$ 0.134 & 1.012 $\pm$ 0.107 & 22.837 $\pm$ 0.147 & 1.079 $\pm$ 0.117 & 0.111 &   86 \\ 
FCC 249 & 13.6 &  23.913 $\pm$ 0.089 & 0.929 $\pm$ 0.068 & 22.935 $\pm$ 0.091 & 0.939 $\pm$ 0.070 & 0.078 &  155 \\ 
FCC 255 & 13.7 &  23.737 $\pm$ 0.111 & 0.780 $\pm$ 0.089 & 22.714 $\pm$ 0.110 & 0.770 $\pm$ 0.087 & 0.125 &   80 \\ 
FCC 277 & 13.8 &  24.244 $\pm$ 0.158 & 0.677 $\pm$ 0.136 & 23.278 $\pm$ 0.156 & 0.683 $\pm$ 0.121 & 0.199 &   42 \\ 
FCC 55  & 13.9 &  24.446 $\pm$ 0.148 & 0.655 $\pm$ 0.111 & 23.441 $\pm$ 0.181 & 0.734 $\pm$ 0.135 & 0.223 &   37 \\ 
FCC 152 & 14.1 &  23.485 $\pm$ 0.344 & 0.844 $\pm$ 0.234 & 22.492 $\pm$ 0.248 & 0.585 $\pm$ 0.179 & 0.456 &   16 \\ 
FCC 301 & 14.2 &  24.383 $\pm$ 0.415 & 1.008 $\pm$ 0.289 & 23.605 $\pm$ 0.628 & 1.216 $\pm$ 0.406 & 0.353 &   21 \\ 
FCC 335 & 14.2 &  23.026 $\pm$ 0.766 & 1.593 $\pm$ 0.585 & 21.954 $\pm$ 0.732 & 1.517 $\pm$ 0.570 & 0.525 &   14 \\ 
FCC 143~*& 14.3 &  23.873 $\pm$ 0.148 & 0.908 $\pm$ 0.114 & 22.929 $\pm$ 0.141 & 0.855 $\pm$ 0.119 & 0.158 &  62 \\ 
FCC 95  & 14.6 &  24.154 $\pm$ 0.098 & 0.263 $\pm$ 0.074 & 23.069 $\pm$ 0.057 & 0.155 $\pm$ 0.046 & 0.373 &   21 \\ 
FCC 136 & 14.8 &  23.968 $\pm$ 0.163 & 0.436 $\pm$ 0.245 & 23.011 $\pm$ 0.104 & 0.355 $\pm$ 0.088 & 0.294 &   25 \\ 
FCC 182 & 14.9 &  24.169 $\pm$ 0.142 & 0.891 $\pm$ 0.111 & 23.220 $\pm$ 0.178 & 1.008 $\pm$ 0.136 & 0.145 &   59 \\ 
FCC 204 & 14.9 &  24.192 $\pm$ 0.518 & 0.944 $\pm$ 0.392 & 23.599 $\pm$ 0.757 & 1.118 $\pm$ 0.498 & 0.443 &   17 \\ 
FCC 119 & 15.0 &  25.464 $\pm$ 0.946 & 0.972 $\pm$ 0.551 & 25.150 $\pm$ 7.387 & 1.222 $\pm$ 0.312 & 0.411 &   17 \\ 
FCC 26  & 15.0 &  23.208 $\pm$ 0.143 & 0.441 $\pm$ 0.114 & 22.394 $\pm$ 0.226 & 0.657 $\pm$ 0.159 & 0.337 &   22 \\ 
FCC 90  & 15.0 &  23.953 $\pm$ 0.352 & 0.673 $\pm$ 0.299 & 23.026 $\pm$ 0.213 & 0.567 $\pm$ 0.211 & 0.370 &   21 \\ 
FCC 106 & 15.1 &  23.966 $\pm$ 1.321 & 1.990 $\pm$ 0.917 & 23.235 $\pm$ 1.879 & 2.223 $\pm$ 1.326 & 0.486 &   15 \\ 
FCC 19  & 15.2 &  24.552 $\pm$ 0.459 & 0.750 $\pm$ 0.291 & 23.483 $\pm$ 0.496 & 0.720 $\pm$ 0.320 & 0.463 &   16 \\ 
FCC 288 & 15.4 &  24.913 $\pm$ 0.510 & 0.871 $\pm$ 0.353 & 24.355 $\pm$ 0.706 & 0.893 $\pm$ 0.682 & 0.426 &   17 \\ 
FCC 202~*& 15.3 &  23.996 $\pm$ 0.084 & 1.101 $\pm$ 0.068 & 22.834 $\pm$ 0.083 & 1.087 $\pm$ 0.069 & 0.050 & 232 \\ 
FCC 324 & 15.3 &  23.698 $\pm$ 0.271 & 0.665 $\pm$ 0.207 & 22.947 $\pm$ 0.274 & 0.758 $\pm$ 0.192 & 0.384 &   21 \\ 
FCC 100 & 15.5 &  24.119 $\pm$ 0.114 & 0.433 $\pm$ 0.098 & 23.366 $\pm$ 0.181 & 0.590 $\pm$ 0.137 & 0.272 &   34 \\ 
FCC 203 & 15.5 &  24.155 $\pm$ 0.361 & 1.184 $\pm$ 0.270 & 23.167 $\pm$ 0.329 & 1.114 $\pm$ 0.241 & 0.271 &   30 \\ 
FCC 303 & 15.5 &  23.479 $\pm$ 0.200 & 0.623 $\pm$ 0.141 & 22.531 $\pm$ 0.218 & 0.678 $\pm$ 0.152 & 0.350 &   22 \\ 
\enddata
\tablenotetext{}{Notes -- (1) Galaxy FCC number. (2) Galaxy $B$-band magnitude. (3) and (4) Maximum likelihood 
estimates of the Gaussian mean $\mu$ and dispersion $\sigma$ of the $g$-band GCLF. (5) and (6) Same as cols.~(3) 
and (4), but for the $z$-band. (7) Fraction of the sample that is expected to be contamination. (8) Total number 
N of all objects (including contaminants and uncorrected for incompleteness) with $p_{GC} \geq 0.5$. (*) These 
galaxies were excluded from the analysis because of their close proximity to massive elliptical galaxies.} 
\label{tab:acsfcs_gclf_par}                                                                                       
\end{deluxetable*}

\begin{deluxetable}{lcc}
\tablecolumns{3}
\tablewidth{0pc}
\tablecaption{Literature compilation of relative distance modulus between Virgo and Fornax clusters}
\tablehead{
\colhead{Method} & \colhead{$\Delta(m-M)$} & \colhead{Reference}}
\startdata
Cepheids          & $0.47 \pm 0.20$ & 1    \\ [0.2ex]
Fund.~Plane       & $0.45 \pm 0.15$ & 2    \\ 
                  & $0.52 \pm 0.17$ & 3    \\ [0.2ex]
PNLF              & $0.35 \pm 0.21$ & 4, 5 \\ 
                  & $0.30 \pm 0.10$ & 6    \\ [0.2ex]
GCLF              & $0.08 \pm 0.09$ & 7    \\
                  & $0.13 \pm 0.11$ & 8    \\
                  & $0.09 \pm 0.27$ & 6    \\
                  & $0.17 \pm 0.28$ & 9    \\ [0.2ex]
SBF               & $0.42 \pm 0.03$ & 10   \\
\enddata
\tablenotetext{}{Notes -- The cited references are: (1) Freedman et~al.~2001, (2) D'Onofrio et~al.~1997, 
(3) Kelson et~al.~2000, (4) Ciardullo et~al.~1998, (5) McMillan et~al.~1993, (6) Ferrarese et~al.~2000a,
(7) Kohle et~al.~1996, (8) Blakeslee \& Tonry 1996, (9) Richtler 2003, (10) Blakeslee et~al.~2009.}
\label{tab:dists}     
\end{deluxetable}


\begin{thebibliography}{} 

\bibitem[Ashman(1995)]{ashman95} Ashman, K.M., Conti, A., \& Zepf, S.E., 
   1995, \aj, 110, 1164
\bibitem[Beasley(2000)]{Beasley00} Beasley, M.A., Sharples, R.M., Bridges, 
   T.J., Hanes, D.A., Zepf, S.E., Ashman, K.M., Geisler, D., 2000, MNRAS, 
   318, 1249
\bibitem[Binggeli(1985)]{virgo_cat} Binggeli, B., Sandage, A., \& Tammann, 
   G.A., 1985, AJ, 90, 1681
\bibitem[Blakeslee(1996)]{env_GCLF} Blakeslee, J.P. \& Tonry, J.L., 1996, 
   \apj, 465, 19
\bibitem[Blakeslee(1999)]{SBFreview} Blakeslee, J.P., Ajhar, E.A., \& Tonry, 
   J.L., 1999, in Post-Hipparcos Cosmic Candles, ed. A. Heck \& F. Caputo 
   (Boston: Kluwer), 181
\bibitem[Blakeslee(2009)]{SBF_Fornax} Blakeslee, J.P., Jord\'an, A., Mei, S., 
   C\^t\'e, P., Ferrarese, L., Infante, L., Peng, E.W., Tonry, J.L., West, 
   M.J., 2009, \apj, 694, 556
\bibitem[Bohringer(1994)]{Virgo_mass} B\"ohringer, H., Briel, U. G., Schwarz, 
   R. A., Voges, W., Hartner, G., \& Trumper, J. 1994, Nature, 368, 828
\bibitem[Bruzual(2003)]{Bruzual} Bruzual, G. \& Charlot, S., 2003, \mnras, 
   344, 1000
\bibitem[Brodie \& Strader(2006)]{brodie_review} Brodie, J.P. \& Strader, J., 
   2006, ARA\&A, 44, 193
\bibitem[Ciardullo(1998)]{PNLF_V} Ciardullo, R., Jacoby, G.H., Feldmeier, 
   J.J., Bartlett, R.E, 1998, \apj, 492, 62
\bibitem[Cote(2004)]{acsvcs_i} C\^ot\'e, P., Blakeslee, J.P., Ferrarese, L., 
   Jord\'an, A., Mei, S., Merritt, D., Milosavljevic, M., Peng, E.W., Tonry, 
   J.L., \& West, M.J., 2004, \apjs, 153, 223 
\bibitem[Cote(2006)]{acsvcs_viii} C\^ot\'e, P., Piatek, S., Ferrarese, L., 
   Jord\'an, A., Merritt, D., Peng, E.W., Ha\c{s}egan, M., Blakeslee, J.P., 
   Mei, S., West, M.J., Milosavljević, M., Tonry, J.L., 2006, ApJS, 165, 57
\bibitem[Cote(2007)]{acsfcs_ii} C\^ot\'e, P., Ferrarese, L., Jord\'an, A., 
   Blakeslee, J.P., Chen, C.W., Infante, L., Merritt, D., Mei, S., Peng, E.W., 
   Tonry, J.L. 2007, \apj, 671, 1456 
\bibitem[DeLucia(2006)]{Mill_early.type} De Lucia, G., Springel, V., White, 
   S.D.M., Croton, D., Kauffmann, G., 2006, MNRAS, 366, 499
\bibitem[Djorgovski(1987)]{FP} Djorgovski, S., Davis, M., 1987, \apj, 313, 59
\bibitem[D'Onofrio(1997)]{FP-D}D'Onofrio, M., Capaccioli, M., Zaggia, S.R., 
   Caon, N., 1997, \mnras, 289, 847
\bibitem[Dressler(1987)]{D} Dressler, A., Lynden-Bell, D., Burstein, D., 
   Davies, R.L., Faber, S.M., Terlevich, R., Wegner, G., 1987, \apj, 313, 42
\bibitem[Drinkwater(2001)]{Fornax_mass} Drinkwater, M.J., Gregg, M.D., \& 
   Colless, M., 2001, \apj, 548, L139 
\bibitem[Durrell(1996)]{dwarfs_fainter} Durrell, P.R., Harris, W.E., Geisler, 
   D., \& Pudritz, R.E., 1996, \aj, 112, 972
\bibitem[Elmegreen(1997)]{E_and_E} Elmegreen, B.G., \& Efremov, Y.N., 1997, 
   \apj, 480, 235 
\bibitem[Fall(1977)]{fall1} Fall, S. M. \& Rees, M. J., 1977, \mnras, 181, 37
\bibitem[Fall(1985)]{fall2} Fall, S. M. \& Rees, M. J., 1985, \apj, 298, 18
\bibitem[Fall(2001)]{Fall_and_Zhang} Fall, S.M., Zhang, Q., 2001, \apj, 561, 
   751
\bibitem[Ferguson(1989a)]{fornax_cat} Ferguson, H. C. 1989a, AJ, 98, 367
\bibitem[Ferguson(1989b)]{fornax_dens}  Ferguson, H. C. 1989b, Ap\&SS, 157, 
   227 
\bibitem[Ferrarese(2000a)]{dist_pap} Ferrarese, L., Mould, J.R., Kennicutt, 
   R.C., Huchra, J., Ford H.C., Freedman, W.L., Stetson, P.B., Madore, B.F., 
   Sakai, S., Gibson, B.K., Graham, J.A., Hughes, S.M., Illingworth, G.D., 
   Kelson, D.D., Macri, L., Sebo, K., Silbermann, N.A. 2000a, \apj, 529, 745
\bibitem[Ferrarese(2000b)]{dist_rev} Ferrarese, L., Ford, H.C., Huchra, J., 
   Kennicutt, R.C., Mould, J.R., Sakai, S., Freedman, W.L., Stetson, P.B., 
   Madore, B.F., Gibson, B.K., Graham, J.A., Hughes, S.M., Illingworth, G.D., 
   Kelson, D.D., Macri, L., Sebo, K., Silbermann, N.A., 2000b, \apjs, 128, 431
\bibitem[Ferrarese(2006)]{Virgo_mz} Ferrarese, L., C\^ot\'e, P., Jord\'an, 
   A., Peng, E.W., Blakeslee, J.P., Piatek, S., Mei, S., Merritt, D., 
   Milosavljevi\'c, M., Tonry, J.L., West, M.J., 2006, \apjs, 164, 334	
\bibitem[Fioc(1997)]{PEGASE} Fioc, M. \& Rocca-Volmerange, B., 1997, A\&A, 
   326, 950
\bibitem[Fouque(2001)]{Virgo_mass} Fouqu\'e, P., Solanes, J. M., Sanchis, 
   T., \& Balkowski, C. 2001, A\&A, 
   375, 770 
\bibitem[Freedman(2001)]{cepheids} Freedman, W.L., Madore, B.F., Gibson, 
   B.K., Ferrarese, L., Kelson, D.D., Sakai, S., Mould, J.R., Kennicutt, 
   R.C., Ford, H.C., Graham, J.A., Huchra, J.P., Hughes, S.M.G., Illingworth, 
   G.D., Macri, L.M., Stetson, P.B., 2001, ApJ, 553, 47
\bibitem[Gavazzi(2005)]{dichotomy} Gavazzi, G., Donati, A., Cucciati, O., 
   Sabatini, S., Boselli, A., Davies, J., Zibetti, S., 2005, A\&A, 430, 411
\bibitem[Gieles(2006)]{Mark} Gieles, M., Larsen, S.S., Scheepmaker, R.A., 
   Bastian, N., Haas, M.R., Lamers, H.J.G.L.M., 2006, A\&A, 446, 9
\bibitem[Graham(2003)]{dichotomy} Graham, A.W., \& Guzm\'an, R., 2003, AJ, 
   125, 2936 
\bibitem[Gnedin(1997)]{gnedin} Gnedin, O.Y. \& Ostriker, J.P., 1997, ApJ, 474, 
   223
\bibitem[Goudfrooij(2001)]{ngc1316} Goudfrooij, P., Alonso, M.V., Maraston, C. 
   \& Minniti, D., 2001, MNRAS, 328, 237
\bibitem[Harris(1994)]{Harris_and_Pudritz} Harris, W.E., \& Pudritz, R.E., 
   1994, \apj, 429, 177 
\bibitem[Harris(2001)]{harris01} Harris, W.E., 2001, in Star Clusters, ed. L. 
   Labhardt \& B. Binggeli (Berlin: Springer), 223 
\bibitem[Harris(2009)]{harris09} Harris, W.E., Kavelaars, J.J., Hanes, D.A., 
   Pritchet, C.J., Baum, W.A., 2009, \aj, 137, 3314 
\bibitem[Izenman(1991)]{izenman91} Izenman, A.J., 1991, J. Am. Stat. Assoc., 
   86, 205
\bibitem[Jacoby(1990)]{PNLF_V} Jacoby, G.H., Ciardullo, R., Ford, H.C., 1990, 
   \apj, 356, 332 
\bibitem[Jacoby et~al.(1992)]{dist_ind_rev} Jacoby, G.H., Branch, D., 
   Ciardullo, R., Davies, R.L., Harris, W.E., Pierce, M.J., Pritchet, C.J., 
   Tonry, J.L., Welch, D.L., 1992, PASP, 104, 599
\bibitem[Jordan(2002)]{color} Jord\'an, A., C\^ot\'e, P., West, M.J. \& Marzke,
   R.O., 2002, \apj, 576L, 113
\bibitem[Jordan(2004)]{acsvcs_red} Jord\'an, A., Blakeslee, J.P., Peng, E.W.,
   Mei, S., C\^ot\'e, P., Ferrarese, L., Tonry, J.L., Merritt, D., 
   Milosavljevic, M., \& West, M.J. 2004, \apjs, 154, 509
\bibitem[Jordan(2006)]{gclf_lett} Jord\'an, A., McLaughlin, D.E., C\^ot\'e, P.,
  Ferrarese, L., Peng, E.W., Blakeslee, J.P., Mei. S., Villegas, D., Merritt, 
  D., Tonry, J.L., \& West, M.J. 2006, \apj, 651, L25.
\bibitem[Jordan(2007a)]{acsfcs_i} Jord\'an, A., Blakeslee, J.P., C\^ot\'e, P., 
  Ferrarese, L., Infante, L., Mei, S., Merritt, D., Peng, E.W., Tonry, J.L. \&
  West, M.J., 2007a, \apjs, 169, 213
\bibitem[Jordan(2007b)]{gclf} Jord\'an, A., McLaughlin, D.E.,
  C\^ot\'e, P., Ferrarese, L., Peng, E.W., Mei. S., Villegas, D.,
  Merritt, D., Tonry, J.L., \& West, M.J. 2007b, \apjs, 171, 101
\bibitem[Jordan(2009)]{vcs_gclf_cats} Jord\'an, A., Peng, E.W.,
   Blakeslee, J.~P., C\^ot\'e, P., Eyheramendy, S., Ferrarese, L., Mei, S., 
   Tonry, J.L., West, M.J., 2009, \apjs, 180, 54
\bibitem[Kelson(2000)]{HST_Key_FP} Kelson, D.D., Illingworth, G.D., Tonry, 
   J.L., Freedman, W.L., Kennicutt, R.C., Mould, J.R., Graham, J.A., Huchra, 
   J.P., Macri, L.M., Madore, B.F., Ferrarese, L., Gibson, B.K., Sakai, S., 
   Stetson, P.B., Ajhar, E.A., Blakeslee, J.P., Dressler, A., Ford, H.C., 
   Hughes, S.M.G., Sebo, K.M., Silbermann, N.A., 2000, \apj, 529, 768
\bibitem[King(1966)]{king66} King, I.R., 1966, \aj, 71, 64
\bibitem[Kissler(1994)]{KP94} Kissler, M., Richtler, T., Held, E.V., Grebel, 
   E.K., Wagner, S.J., Capaccioli, M., 1994, 1994, A\&A, 287, 463 
\bibitem[Kohle(1996)]{KP96} Kohle, S., Kissler-Patig, M., Hilker, M., 
Richtler, T., Infante, L., Quintana, H., 1996, A\&A, 309, 39 
\bibitem[Kormendy(1985)]{dichotomy} Kormendy, J. 1985, ApJ, 295, 73
\bibitem[Kundu(2001)]{gclf_measurements} Kundu, A., Whitmore, B.C., 2001, 
   2001, \aj, 121, 2950
\bibitem[Lamers(2006)]{mass_loss_rate} Lamers, H.J.G.L.M. \& Gieles, M. 2006, 
   A\&A, 455, L17
\bibitem[McLaughlin(1999)]{mclaughlin99} McLaughlin, D. E., 1999, AJ, 117, 
   2398 
\bibitem[McLaughlin(2008)]{ManfF} McLaughlin, D.E. \& Fall, S.M., 2008, \apj, 
   679, 1272
\bibitem[McMillan(1993)]{PNLF_f} McMillan, R., Ciardullo, R., Jacoby, G.H., 
   1993, \apj, 416, 62
\bibitem[Mei(2005a)]{mei05a} Mei, S., Blakeslee, J.P., Tonry, J.L., Jord\'an, 
   A., Peng, E.W., C\^ot\'e, P., Ferrarese, L., Merritt, D., Milosavljevic, 
   M., \& West, M.J., 2005a, \apjs, 156, 113.
\bibitem[Mei(2005b)]{mei05b} Mei, S., Blakeslee, J.P., Tonry, J.L., Jord\'an, 
   A., Peng, E.W., C\^ot\'e, P., Ferrarese, L., West, M.J., Merritt, D., \& 
   Milosavljevic, M., 2005b, \apj, 625, 121
\bibitem[Mei(2007)]{mei07} Mei, S., Blakeslee, J.P., C\^ot\'e, P., Tonry, 
   J.L., West, M.J., Ferrarese, L., Jord\'an, A., Peng, E.W., Anthony, A., 
   \& Merritt, D., 2007, \apj, 655, 144
\bibitem[Miller(2007)]{dwarf_faint} Miller, B.W. \& Lotz, J.M., 2007, \apj, 
   670, 1074
\bibitem[Peebles(1968)]{Peebles68} Peebles, P.J.E. \& Dicke, R.H., 1968, 
   \apj, 154, 891
\bibitem[Peng(2006a)]{peng06a} Peng, E.W., Jord\'an, A., C\^ot\'e, P., 
   Blakeslee, J.P., Ferrarese, L., Mei, S., West, M.J., Merritt, D., 
   Milosavljevic, M., \& Tonry, J.L., 2006, \apj, 639, 95
\bibitem[Prieto(2006)]{jose} Prieto, J.L. \& Gnedin, O.Y., 2008, \apj, 689, 
   919
\bibitem[Puzia(1999)]{Putzia99} Puzia, T.H., Kissler-Patig, M., Brodie, J.P., 
   \& Huchra J.P., 1999, AJ, 118, 2734
\bibitem[Puzia(2002)]{Putzia99} Puzia, T.H., Zepf, S.E., Kissler-Patig, M., 
    Hilker, M., Minniti, D., Goudfrooij, P., A\&A, 391, 453
\bibitem[Richtler(2003)]{richtler} Richtler T., 2003, LNP Vol. 635: Stellar 
    Candles for the Extragalactic Distance Scale, 635, 281
\bibitem[Salpeter(1955)]{imf} Salpeter, E.E., 1955, \apj, 266, 713
\bibitem[red-map]{Schelegel} Schlegel, D.J., Finkbeiner, D.P., \& Davis, M., 
   1998, \apj, 500, 525
\bibitem[Secker(1992)]{secker} Secker, J., 1992, AJ, 104, 1472 
\bibitem[secker96]{GCLF fitting model} Secker, J. \& Harris, W. E., 1993, 
   \aj, 105, 1358
\bibitem[Schindler(1999)]{virgo_mass} Schindler, S., Binggeli, B., \& 
   B\"ohringer, H. 1999, A\&A, 343, 420 
\bibitem[Springel2005]{milenio} Springel, V., White, S.D.M., Jenkins, A., 
   Frenk, C.S., Yoshida, N., Gao, L., Navarro, J., Thacker, R., Croton, D., 
   Helly, J., Peacock, J.A., Cole, S., Thomas, P., Couchman, H., Evrard, A., 
   Colberg, J., Pearce, F., 2005, Nature, 435, 629
\bibitem[Tonry(1988)]{sbf} Tonry, J. L., \& Schneider, D. P. 1988, AJ, 96, 
   807
\bibitem[Tonry(2000)]{virgo_mass} Tonry, J. L., Blakeslee, J. P., Ajhar, 
   E. A., \& Dressler, A. 2000, ApJ, 530, 625
\bibitem[Schlegel(1998)]{schlegel98} Schlegel, D.J., Finkbeiner, D.P. \& 
   Davis, M., 1998, \apj, 500, 525
\bibitem[Vesperini(2000)]{vesperini1} Vesperini, E., 2000, MNRAS, 318, 841
\bibitem[Vesperini(2001)]{vesperini2} Vesperini, E., 2001, MNRAS, 322, 247
\bibitem[West(1993)]{west} West, M.J., 1993, \mnras, 265, 755
\bibitem[Whitmore(1997)]{whit} Whitmore, B. C. 1997, in The Extragalactic 
   Distance Scale, ed. M. Livio, M. Donahue, \& N. Panagia (Baltimore: STScI), 
   254
\end{thebibliography}
\end{document}